\documentclass[sigconf]{acmart}

\usepackage{natbib}
\AtBeginDocument{%
  \providecommand\BibTeX{{%
    \normalfont B\kern-0.5em{\scshape i\kern-0.25em b}\kern-0.8em\TeX}}}

\newcommand{\wlrevised}[1]{\textcolor{black}{#1}}
\newcommand{\ycrevised}[1]{\textcolor{black}{#1}}

\copyrightyear{2024}
\acmYear{2024}
\setcopyright{acmlicensed}\acmConference[CHI '24]{Proceedings of the CHI Conference on Human Factors in Computing Systems}{May 11--16, 2024}{Honolulu, HI, USA}
\acmBooktitle{Proceedings of the CHI Conference on Human Factors in Computing Systems (CHI '24), May 11--16, 2024, Honolulu, HI, USA}
\acmDOI{10.1145/3613904.3642800}
\acmISBN{979-8-4007-0330-0/24/05}

\begin{document}

\title{Exploring the Design of Generative AI in Supporting Music-based Reminiscence for Older Adults}

\author{Yucheng Jin}
\authornote{Both authors contributed equally to this research.}
\affiliation{%
  \institution{Department of Computer Science, Hong Kong Baptist University}
  \city{Hong Kong}
  \country{China}
}
\email{yuchengjin@hkbu.edu.hk}

\author{Wanling Cai}
\authornotemark[1]
\affiliation{%
  \institution{Lero @ Trinity College Dublin}
  \city{Dublin}
  \country{Ireland}
}
\email{wanling.cai@tcd.ie}

\author{Li Chen}
\affiliation{%
  \institution{Department of Computer Science, Hong Kong Baptist University}
  \city{Hong Kong}
  \country{China}
}
\email{lichen@comp.hkbu.edu.hk}

\author{Yizhe Zhang}
\affiliation{%
  \institution{Department of Computer Science, Hong Kong Baptist University}
  \city{Hong Kong}
  \country{China}
}
\email{csyzzhang@comp.hkbu.edu.hk}

\author{Gavin Doherty}
\affiliation{%
  \institution{School of Computer Science and Statistics, Trinity College Dublin}
  \city{Dublin}
  \country{Ireland}
}
\email{gavin.doherty@tcd.ie}

\author{Tonglin Jiang}
\affiliation{%
  \institution{School of Psychological and Cognitive Sciences, Peking University}
  \city{Beijing}
  \country{China}
}
\email{tljiang@pku.edu.cn}

\renewcommand{\shortauthors}{Jin et al.}

\newcommand{\romannum}[1]{\uppercase\expandafter{\romannumeral#1}}
\newcommand{\circlenum}[1]{\raisebox{.5pt}{\textcircled{\raisebox{-.9pt} {#1}}}}
\begin{abstract}
Music-based reminiscence has the potential to positively impact the psychological well-being of older adults. However, the aging process and physiological changes, such as memory decline and limited verbal communication, may impede the ability of older adults to recall their memories and life experiences. Given the advanced capabilities of generative artificial intelligence (AI) systems, such as generated conversations and images, and their potential to facilitate the reminiscing process, this study aims to explore the design of generative AI to support music-based reminiscence in older adults. This study follows a user-centered design approach incorporating various stages, including detailed interviews with two social workers and two design workshops (involving ten older adults). Our work contributes to an in-depth understanding of older adults’ attitudes toward utilizing generative AI for supporting music-based reminiscence and identifies concrete design considerations for the future design of generative AI to enhance the reminiscence experience of older adults. 

\end{abstract}




\maketitle

\section{Introduction}
Reminiscence involves the act of recalling and reflecting on past memories and personal experiences~\cite{woods1992reminiscence}, which has been demonstrated to offer numerous benefits for older adults~\cite{bohlmeijer2007effects,cappeliez2008functions, hallford2013reminiscence}. For example, reminiscing about positive memories and experiences can improve mood and reduce feelings of loneliness, depression, and anxiety that older adults may likely experience~\cite{bryant2005using,
cappeliez2008functions}. Engaging in reminiscence activities can also enhance cognitive functions such as memory and attention~
\cite{cotelli2012reminiscence}, which significantly contributes to successful aging~\cite{bohlmeijer2007effects}. However, reminiscence can be challenging for older adults due to memory decline, cognitive impairments, and articulation~\cite{lin2003effect}. Prior work has suggested that reminiscence can be prompted through various types of stimuli, such as old photographs, familiar songs, specific foods, or even particular scents~\cite{romaniuk1981looking}. Among them, music, deeply connected with emotions and effective in triggering memories~\cite{schafer2013psychological}, has been increasingly used for engaging older adults in reminiscence and enhancing their experience and has been shown to have positive effects on their psychological well-being~\cite{janata2007characterisation,dassa2018musical}.

In recent decades, HCI researchers have investigated the use of technology for supporting reminiscence~\cite{caprani2005remember,petrelli2008autotopography,tsai2013framing,yoo2021understanding,lazar2014systematic}, and different types of technology-based tools have been designed to enable people to re-experience their personal digital data like social media content~\cite{peesapati2010pensieve}, emails~\cite{hangal2011muse},  photos~\cite{petrelli2008autotopography,axtell2019photoflow,czech2020discovering}, sound~\cite{hsieh2011soundcapsule,jayaratne2016memory} and music~\cite{sas2020supporting}, to trigger their memories. Moreover, several researchers have proposed AI-driven solutions like chatbots that aim to offer human-like and personalized interactions~\cite{nikitina2018smart,caros2020automatic,gamborino2021towards}, for realizing the potential of AI technologies to aid the reminiscing process. 

With the recent advent of generative AI  and its increasing capabilities for creating new content such as text, images, and music~\cite{zhang2023complete}, we think that, in the context of music reminiscence, it might be a potentially powerful tool to enhance older adults' reminiscence experiences. For example, AI may generate either questions or images to help older adults reminisce about specific life events in greater detail. However, limited HCI research has investigated this topic so far.

Therefore, in this study, we are interested in exploring the design of generative AI for older adults to enrich their experience of music reminiscence. Given the importance of accessibility and usability in the design of reminiscence technology for older adults~\cite{petrelli2008autotopography,lazar2014systematic}, it is crucial to first understand the perspectives of older adults on generative AI technologies in the context of music reminiscence, and then examine how generative AI could be designed to support their music-based reminiscing activities. Thus, we have raised two research questions in our study: 

\textbf{RQ1:} \textit{What are older adults' perspectives on generative AI for music reminiscence?}

\textbf{RQ2:} \textit{What are the considerations that should be taken into account for designing generative AI to support music reminiscence among older adults?}


To address these two questions, \ycrevised{we contacted two social workers who work in a local community center. They helped recruit seven older adults for Workshop \romannum{1} and nine older adults for Workshop \romannum{2} (with six participants attending both workshops) in Hong Kong, China.} Specifically, we first conducted a group interview with the two social workers, which sought to understand older adults' daily practice with emerging computer technologies and discuss the potential use of generative AI (i.e., AI-generated conversations and images) to facilitate the process of music reminiscence. Based on the insights from the interviews, we prepared our study materials and conducted two design workshops with older adults. In \wlrevised{Workshop \romannum{1}}, we showed our initial design concepts using a video prototype and conducted a semi-structured group interview to gain an in-depth understanding of older adults' perspectives on generative AI in the context of music reminiscence, based on which we designed a digital prototype of a generative AI based reminiscence tool. In \wlrevised{Workshop \romannum{2}}, we evaluated the digital prototype with \wlrevised{nine older adults} and conducted a follow-up group interview to examine the appropriate design of generative AI for supporting music reminiscence among older adults. 

Through reflexive thematic analysis of data from the two design workshops, we found that \ycrevised{the older adults of our study} perceived AI-generated conversations and images to be useful in their reminiscence process, as such content might help them recall more details of their life experiences and also sustain their engagement. Moreover, when using generative AI systems to support music reminiscence, most older adults expressed a preference for individual interactions with AI systems rather than interaction in group settings, as they think people may have different feelings when listening to a certain song and group sharing could potentially weaken their personal reminiscence experiences. With regards to the design of generative AI for reminiscence, older adults expected that AI-generated content, including conversations and images, can be more personalized and relevant to their own memories. They also noted that the quality and relevance of AI-generated content would highly influence their reminiscence experiences and their willingness to use such technologies. In addition, they showed a desire for autonomy when using generative AI systems for reminiscence; for example, they would expect control over input modalities (e.g., voice and text) and image generation (e.g., to be able to select and refine the generated images). Finally, based on our findings, we present several design implications for using generative AI to support reminiscence among older adults \ycrevised{in our study}, which will be of interest to researchers involved in reminiscence technology designed for older adults.

\section{Related Work}

\subsection{Benefits of Reminiscence for Older Adults}
Reminiscence, defined as ``the act of recalling and reflecting upon past life experiences and memories, either alone or with a companion or group of people~\cite{woods1992reminiscence}'', has often been viewed as a form of autobiographical memory~\cite{bluck1998reminiscence} or life story work~\cite{butler1963life}. Prior studies in psychology and gerontology have demonstrated various functions and benefits of reminiscence for older adults, including the enhancement of psychological well-being, cognitive function, and overall quality of life~\cite{bryant2005using,lewis1971reminiscing,bohlmeijer2007effects,cappeliez2008functions,hsieh2003effect, hallford2013reminiscence}. For instance, reminiscing about worthwhile experiences from the past may provide older adults a sense of purpose and fulfillment, which can enhance their meaning in life~\cite{wong2013processes}. Reflecting on how they dealt with past difficulties can increase their sense of mastery and self-confidence~\cite{bohlmeijer2007effects}. 

In the past decades, activities such as reminiscence therapy sessions have been designed to facilitate reminiscence in elderly populations~\cite{hsieh2003effect,hallford2013reminiscence}. 
Within reminiscence activities, different items such as photographs and recordings of music are often used as memory triggers to initiate the process of recollection~\cite{romaniuk1981looking}. In particular, music has been increasingly integrated into reminiscence practice~\cite{istvandity2017combining}, because music has been shown to induce psychological benefits in various age groups~\cite{schafer2013psychological}. Previous studies on music-based reminiscence have shown that music can enhance autobiographical recall by promoting the recall of memories \wlrevised{linked to their self-identity and life story, which could enhance the sense of meaning in life and foster self-continuity (i.e., a sense of connection between one's past and one's present)}~\cite{janata2007characterisation,dassa2018musical,engelbrecht2023music, sedikides2022psychological}.

\subsection{Technology for Reminiscence}
In recent decades, there has been increasing attention within the HCI community on designing technology to support and enhance people's reminiscence of their past life experiences~\cite{caprani2005remember,petrelli2008autotopography,peesapati2010pensieve,tsai2013framing,axtell2019photoflow,yoo2021understanding}. Researchers and designers have created various technology-based reminiscence tools that utilize people's digital data, such as social media content~\cite{peesapati2010pensieve}, emails~\cite{hangal2011muse}, and photos~\cite{petrelli2008autotopography,axtell2019photoflow,czech2020discovering}, as resources to stimulate their memories and aid in the process of reminiscing about the past. For example, \citeauthor{peesapati2010pensieve} designed a system, Pensieve~\cite{peesapati2010pensieve}, which encourages people to reminisce and helps people write about their reminiscence by sending memory triggers through emails. These triggers contain social media content that people have created on social media websites, e.g., Flickr and Last.fm, or generic prompts that relate to common life experiences. \citeauthor{hangal2011muse} presented a system, MUSE~\cite{hangal2011muse}, which analyzes the content of long-term email archives and produces a set of prompts (e.g., recurring named entities, image attachments) to trigger users' memories, aiming to assist people in browsing email archives and reminiscing on their past lives. 

Due to the widespread use of smartphones, many people are creating increasingly large personal photography collections. A body of reminiscence work has investigated supporting reminiscence using digital photo collections~\cite{axtell2019photoflow,mcgookin2019reveal}. For instance, \citeauthor{mcgookin2019reveal} presented a location-based reminiscing tool, which leverages changes in user locations to initiate searches of a photo library for images taken nearby~\cite{mcgookin2019reveal}. This location-based approach was shown to encourage more individual reflection on their daily photographic practice and establish a link between the images and the current moment, enhancing the reminiscing experience. Moreover, \citeauthor{axtell2019photoflow} developed PhotoFlow~\cite{axtell2019photoflow} to support older adults and their families in engaging in reminiscence activities (e.g., sharing of family memory) using their family photographs and utilizing their speech data to organize photos automatically. 
In addition to employing visual cues such as text and photos to trigger memories, sensory stimuli such as sound and music have also been used to facilitate the reminiscing process~\cite{hsieh2011soundcapsule,czech2020discovering,jayaratne2016memory,sas2020supporting}. \citeauthor{jayaratne2016memory} explored the use of sound (e.g., audio recording) as a tool to support people's reminiscing about their loved ones \cite{jayaratne2016memory}. \citeauthor{odom2019investigating} designed a music player aimed at supporting people's experience of reflection and reminiscence by enabling them to re-experience the music they enjoyed in the past~\cite{odom2019investigating}. 
In short, existing studies have investigated the use of personal digital data and interactive technologies to support reminiscence. 
Benefits of these technological solutions have been found in older adults~\cite{axtell2019photoflow,caprani2005remember}, especially people with dementia~\cite{edmeads2019designing, dixon2020approach,czech2020discovering,sas2020supporting,bejan2018virtual} or blindness~\cite{yoo2021understanding,yoo2020understanding}.

\subsection{Generative AI}

Generative Artificial Intelligence (Generative AI) has gained increasing prevalence in recent years due to its remarkable capabilities in creating new content similar to human-generated content~\cite{zhang2023complete}. Generative AI technologies (e.g., based on Large Language Models (LLMs)) and associated tools such as Midjourney and ChatGPT empower people to write creative stories and generate images and videos. The use of these tools has been studied in various domains such as education~\cite{baidoo2023education} and health~\cite{nova2023generative}.  For instance, \citeauthor{han2023design} designed an AI-based visual storytelling application that augments children's creative expression and storytelling and develops their literacy~\cite{han2023design}. \citeauthor{mirowski2023co} designed a system based on LLMs for collaborative script writing by generating coherent theatre scripts and screenplays \cite{mirowski2023co}. In light of the capabilities (e.g., automatic story generation) of generative AI, we believe that, in the context of music reminiscence, generative AI may also serve as a potentially useful tool to enrich the reminiscence experience of older adults. For instance, AI could generate relevant questions and images to assist older adults in recalling specific life events. 

However, to the best of our knowledge, only a few studies have investigated AI-based solutions such as chatbots to support reminiscence~\cite{caros2020automatic,gamborino2021towards}, which, however, have mainly focused on technical solutions and do not consider user perspectives. Building upon prior work, in this study, we seek to investigate older adults' perspectives on generative AI for music reminiscence and further explore the design of generative AI to enhance their reminiscence experiences.

\begin{figure*}[h]
  \centering
  \includegraphics[width=\linewidth]{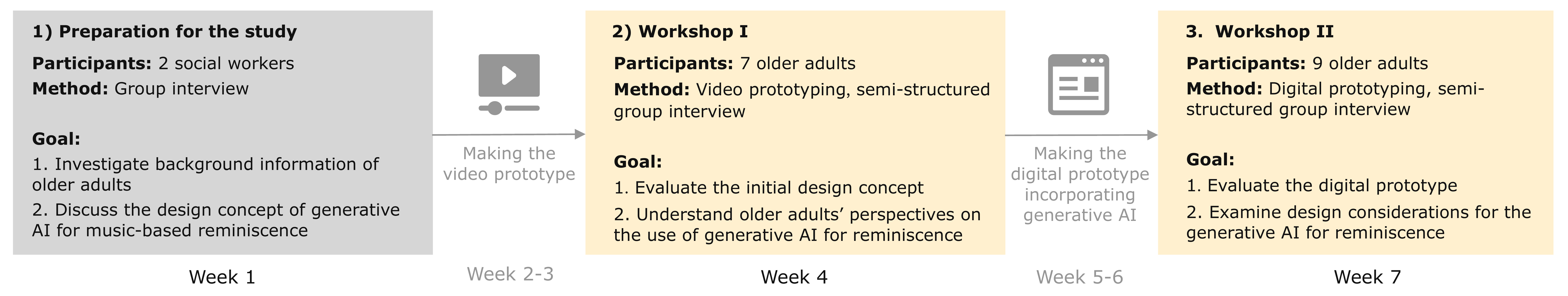}
  \caption{Overall research design and procedure.}
  \label{fig:research_procedure}
  \Description{This flowchart contains the overall study design and process, including one study preparation group interview and two workshops with older adults.}
\end{figure*}

\section{Method}
This study aims to investigate the design of generative AI for older adults in music reminiscence. Specifically, we conducted two design workshops with two goals:
\begin{itemize}
    \item To understand older adults' perspectives on using generative AI for music reminiscence;
    \item To identify design opportunities and challenges of using generative AI to support older adults in music reminiscence.
\end{itemize}

We followed a user-centered design approach and used the workshops to collect participants' feedback on the design prototypes and to discuss using generative AI to perform the activities involved in music reminiscence. Figure~\ref{fig:research_procedure} shows the overall research design and procedure consisting of 
two phases of design and user study (i.e., Workshop \romannum{1} and Workshop \romannum{2}).
\wlrevised{The iterative design approach allows us to better understand older adults' perspectives and integrate their feedback into the design and prototyping.}
We organized the two workshops with ten older adults. We have translated the study materials (e.g., user interfaces and conversation scripts) and the interview quotations into English for this paper.

\subsection{Preparation Work}
We approached a local community center in Hong Kong, China to conduct our study. The center provides developmental and support services to people aged 60 or above and prospective retirees (aged 50-59). To prepare the study materials, we interviewed two social workers who have worked in the center for over five years. One of the social workers specializes in developing music programs for improving the mental well-being of elderly people, such as choir, instrument performance, and music mindfulness. By interviewing the two social workers, we intended to understand their practices for supporting elderly people with these activities and their experiences of working with elderly people. \ycrevised{Based on the interviews with the social workers, several key understandings about older adults in this community center were identified.}

\begin{itemize}
    \item \ycrevised{\textit{Participation in music activities}: The social workers found that some older adults enjoy participating in music activities organized by the center. This is probably because they can cultivate social connections with others, which may boost their confidence and sense of belonging.}
    \item \ycrevised{\textit{Preference for familiar songs}: The social workers observed that, in general, older adults prefer to listen to or play familiar songs. These songs evoke nostalgia and bring back memories, which can be comforting and enjoyable for older adults.}
    \item \ycrevised{\textit{Willingness to learn new technologies}: Contrary to the stereotype that older adults are resistant to technology, the social workers found that most older adults in the center are open to learning new technologies if they believe they could benefit from them.}
    \item \ycrevised{\textit{Digital literacy among older adults}: The social workers noted that some older adults have good digital literacy skills, thanks to the training courses provided by the center. These courses cover various mobile apps like Facebook, YouTube, and WhatsApp.}
\end{itemize}

\ycrevised{
Furthermore, during the interview with the social workers, several potential design considerations for incorporating generative AI in music-based reminiscence were discussed. These considerations aim to enhance the reminiscing experience for older adults. The key points are listed below:
}

\begin{itemize}
    \item \ycrevised{\textit{Interaction conceptualization}: Dialogue-based interaction was identified as a valuable approach to guiding older adults in recalling and sharing their memories evoked by music. This interaction type may facilitate a deeper and more meaningful reminiscing process.}
    \item \ycrevised{\textit{Design concepts}: The interview suggested the importance of designing concepts that align with older adults' mental models of tools for reminiscence. Several design concepts were proposed, including a tree hollow, a diary book, and a smart radio. These concepts aim to create a familiar and comfortable interface that resonates with older adults' existing understanding of reminiscence tools.}
    \item \ycrevised{\textit{Use of generative AI}: We discussed the potential of generative AI in generating texts and questions that facilitate reminiscence. These generated texts could help older adults delve deeper into their memories and emotions. Additionally, generative AI can be used to create images to enrich the details of the memories evoked by the music, making the reminiscing experience more vivid and immersive.}
    \item \ycrevised{\textit{Types of reminiscence activity}: Reminiscence activities can be conducted in either individual or group settings, and older adults may have their own preferences for the type of reminiscence activity they engage in.}

\end{itemize}

\ycrevised{
These findings suggest that older adults in this community center can derive benefits from participating in music activities. They also exhibited preferences for familiar songs and expressed openness to learning new technologies. These insights can guide the development of generative AI systems for music reminiscence tailored to the needs and interests of older adults. Considering the potential advantages of dialogue-based interaction and the discussed design concepts, we opted to create a DJ (referring to a person playing recorded music for an audience) powered by generative AI. To avoid restricting the design of generative AI with a specific product or application in the early design phase, we chose to produce a video prototype to facilitate discussions on how generative AI could support activities related to music reminiscence for older adults.
}

The social workers further helped advertise our study. They asked the participants to finish a pre-registration by answering some questions to acquire their demographics as shown in Table~\ref{demographics} and indicating two songs that could revoke reminiscence along with a brief description of memories evoked by the songs. The collected songs and reminiscence descriptions helped us prepare the content of the prototypes.

\subsection{Participants}
\ycrevised{We recruited participants from the same community center as the preparatory interview with the two social workers.} 
Most of these individuals reside in the district and regularly engage in various social activities offered at the center. Despite obtaining information regarding participants' frequency of music listening and their willingness to listen to nostalgic music, we did not employ any filters based on these characteristics. This decision was driven by the understanding that music reminiscence could benefit older adults in general~\cite{engelbrecht2023music}. 
\ycrevised{The determination of the age range for older adults lacks consensus and can be influenced by cultural and sociodemographic factors~\cite{hedgeman2018perceived}. In our study, we included adults over the age of 55. This aligns with major initiatives such as the National Health Interview Survey~\cite{schoenborn2009health}, in which older adults are categorized into four age groups: 55–64 years, 65–74 years, 75–84 years, and 85 years or above.}

\ycrevised{Seven individuals participated in Workshop \romannum{1}, and nine individuals participated in Workshop \romannum{2}. In total, there were ten distinct participants across the two workshops, with six individuals attending both workshops.}
Of the ten participants, all of them are above 55, five are female, and six hold a bachelor's degree or higher qualification. We compensated each participant with a supermarket coupon valued at 200 HKD (about 25.55 USD) for each workshop.

\begin{table*}[h]
\centering
\caption{Demographics of participants in design workshops of generative AI technologies for music reminiscence}
\label{demographics}
\Description{The table describes the background information and questionnaire results of the ten participants, four were above 65, five were female, and six held a bachelor’s degree or higher qualification. Additionally, six participants participated in two design workshops.}
\begin{tabular}{ccccccc}
\hline

\textbf{ID} & \textbf{\begin{tabular}[c]{@{}l@{}}Age \\ range (y/o)\end{tabular}} & \textbf{\begin{tabular}[c]{@{}l@{}}Gender\end{tabular}} & \textbf{\begin{tabular}[c]{@{}l@{}}Education\\ background\end{tabular}} & \textbf{\begin{tabular}[c]{@{}l@{}}Music listening \\ frequency\end{tabular}} & \textbf{\begin{tabular}[c]{@{}l@{}}Willingness to listen\\ to nostalgic music\end{tabular}} & \textbf{\begin{tabular}[c]{@{}l@{}}Joined \\workshop\end{tabular}}
\\ \hline

P1 & 55-65 & Female & Middle School & Several times per week & Sometimes & \begin{tabular}[c]{@{}l@{}}1\&2\end{tabular}
\\

P2 & 55-65 & Female & Bachelor & Several times per day & Frequently & \begin{tabular}[c]{@{}l@{}} 1\&2 \end{tabular}
\\

P3 & > 65 & Male & Bachelor & Seldom & Sometimes & \begin{tabular}[c]{@{}l@{}} 1 \end{tabular}
\\

P4 & 55-65 & Male & Middle School & Several times per day & Frequently & \begin{tabular}[c]{@{}l@{}} 1\&2 \end{tabular}
\\

P5 & 55-65 & Female & Master & Several times per week & Sometimes & \begin{tabular}[c]{@{}l@{}} 1\&2 \end{tabular}
\\

P6 & 55-65 & Female & Middle School & Several times per day & Frequently & \begin{tabular}[c]{@{}l@{}} 1\&2 \end{tabular}
\\

P7 & 55-65 & Male & Middle School & Several times per week & Sometimes & \begin{tabular}[c]{@{}l@{}} 1\&2 \end{tabular}
\\

P8 & > 65 & Female & Master & Several times per week & Frequently & \begin{tabular}[c]{@{}l@{}} 2 \end{tabular}
\\

P9 & > 65 & Male & Master & Several times per day & Frequently & \begin{tabular}[c]{@{}l@{}} 2 \end{tabular}
\\

P10 & > 65 & Male & Bachelor & Several times per week & Frequently & \begin{tabular}[c]{@{}l@{}} 2 \end{tabular}
\\

\hline
\end{tabular}{}
\end{table*}

\subsection{Workshop \romannum{1}: Understanding Older Adults}

\begin{figure*}[h]
  \centering
  \includegraphics[width=.95\linewidth]{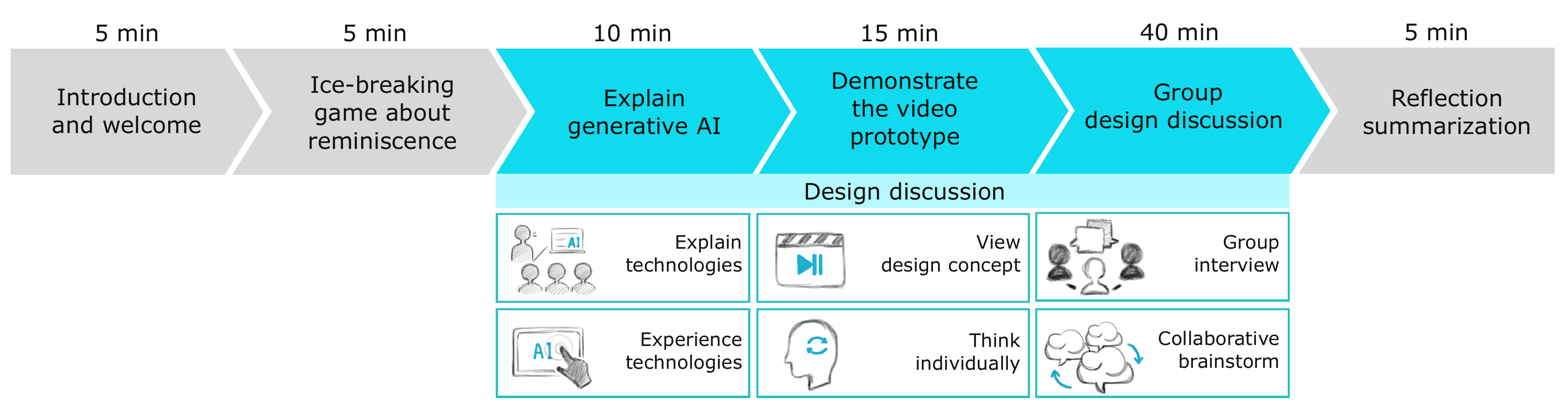}
  \caption{The overview of Workshop \romannum{1} including various activities ranging from explaining generative AI and video prototype demonstration to group design discussion.}
  \label{fig:ws1}
  \Description{This flow chart describes the experimental process of the first Workshop and includes various activities ranging from explaining generative AI and video prototype demonstration to group design discussion.}
\end{figure*}

To evaluate our initial design concept represented by a video prototype (\textit{AI-DJ}) and collect older adults' perspectives on using generative AI in music reminiscence, we designed a set of activities in Workshop \romannum{1} (see Figure~\ref{fig:ws1}). Two facilitators and two collectors worked with seven participants to perform the activities in the workshop. The workshop began with an introduction, followed by an ice-breaking game where we invited participants to share their feelings about viewing some pictures of products used between 1970 and 1990 in Hong Kong. This game helps participants understand the meaning of reminiscence. Then, we conducted various activities with the participating older adults, ranging from explaining generative AI and video prototype demonstration to group design discussion.

\textit{Explain Generative AI.}
Most participants did not have any experience with generative AI. Therefore, we designed two activities to help participants understand generative AI: 1) explaining what they can do with these technologies; and 2) guiding them to experience ChatGPT and MidJourney to generate short stories and pictures related to their interests. 

\begin{figure}[h]
  \centering
  \includegraphics[width=.7\linewidth]{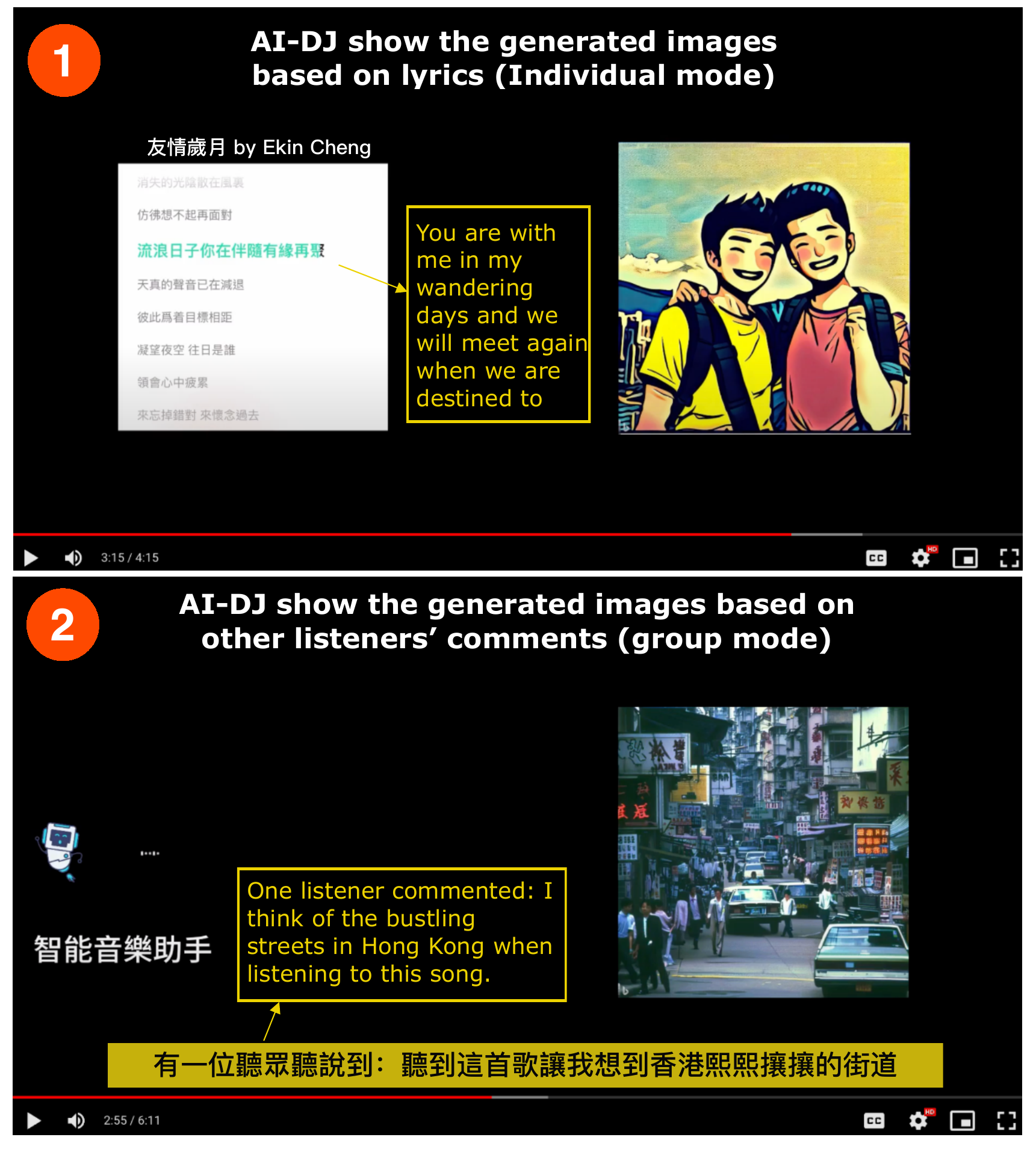}
  \caption{The screenshots of the video prototype (\textit{AI-DJ}) used in Workshop \romannum{1}. (1) \textit{AI-DJ} generated images based on lyrics in the individual setting. (2) \textit{AI-DJ} generated images based on other listeners' comments in the group setting.
}
  \label{fig:ws1_video}
  \Description{This screenshot shows the video prototype (Al-DJ) used in Workshop 1. Subfigure 1 (left) shows Al-DJ generating images based on lyrics in personal mode. Subfigure 2 (right) shows \textit{AI-DJ} generating images based on comments from other listeners in group mode.}
\end{figure}

\textit{Demonstrate Video Prototypes.} To evaluate our initial design concept with the older adults, we created a video prototype based on the discussion with the two social workers in the community center (see Section 3.1). Using a video prototype allows us to discuss design ideas beyond the screen and focus on the use of generative AI in reminiscence contexts. In addition, the video prototype could present design ideas vividly and efficiently and support participatory design~\cite{zwinderman2013using,mackay1999video}. 
Specifically, the video prototype demonstrates how a virtual DJ (\textit{AI-DJ}) guides listeners to \textit{recall} and \textit{share} their memories while listening to nostalgic songs. The video presents two design features: 1) presenting images reflecting the lyrics, music background information, and music memories [Figure~\ref{fig:ws1_video} (1)], and 2) sharing background information about the music and other listeners' memories evoked by the music [Figure~\ref{fig:ws1_video} (2)]. We prepared the dialogue scripts and images using large language models (LLMs), such as BingChat\footnote{\url{https://www.microsoft.com/en-us/edge/features/bing-chat}} and Midjourney\footnote{\url{https://www.midjourney.com/}}. The selection of music is based on the nostalgic songs we collected through the pre-registration form. The video prototype covers two settings for music reminiscence, i.e., \textit{individual} and \textit{group}, because music reminiscence can be either an individual or a group activity~\cite{istvandity2017combining}. Compared with the individual setting, the group setting involves social interaction coordinated by \textit{AI-DJ} where multiple music listeners could share their memories with each other. We asked participants to view the video prototype and then think of their feelings about the design features powered by the generative AI for reminiscence. 

 \begin{figure}[h]
  \centering
  \includegraphics[width=.95\linewidth]{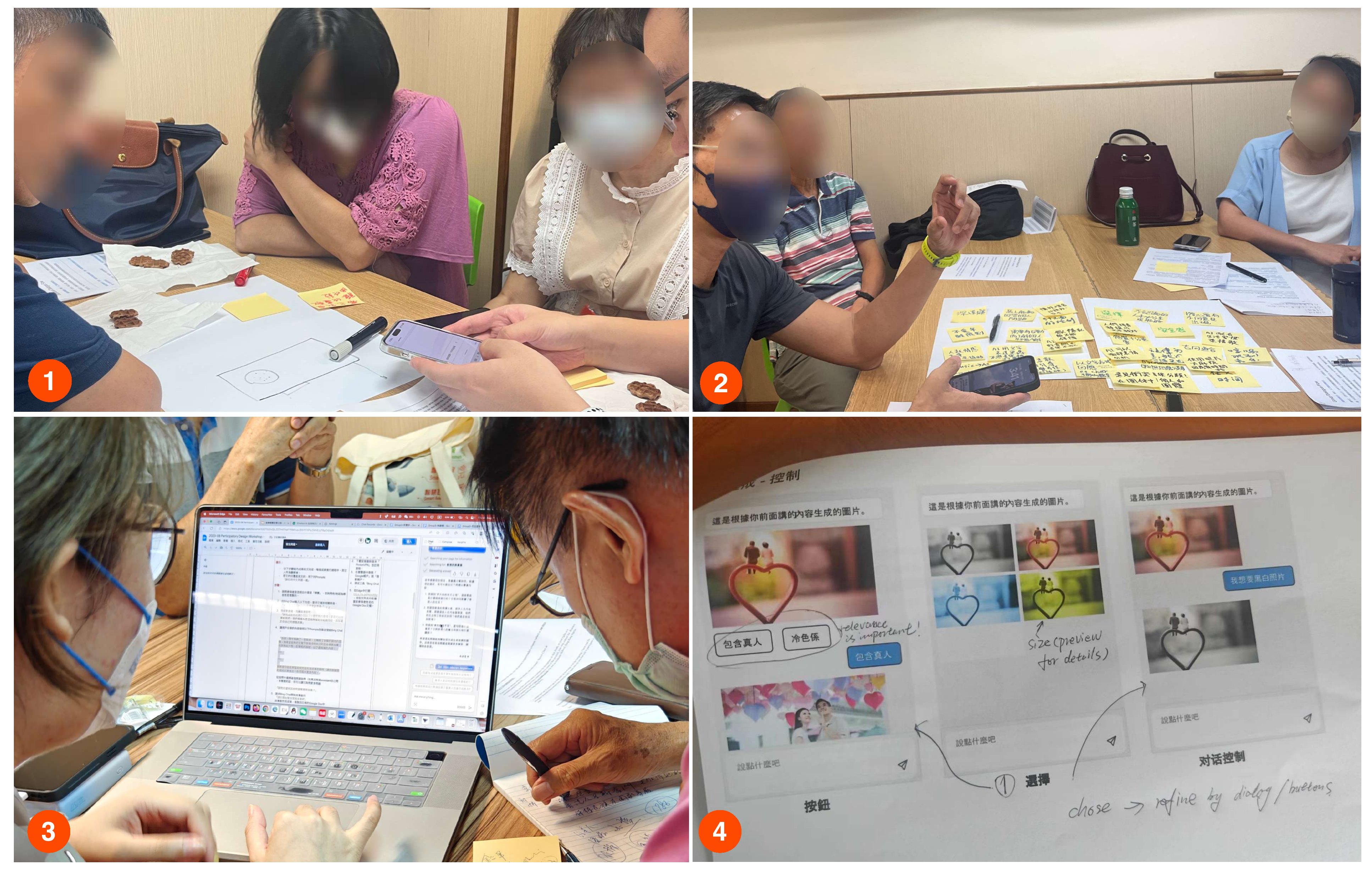}
  \caption{The design activities conducted in the two design workshops. In Workshop \romannum{1}: (1) the participants discussed the design concept represented by the video prototype in a semi-structured group interview; (2) the participants generated new design ideas through collective brainstorming. In Workshop \romannum{2}: (3) the participants interacted with the digital prototype (\textit{MusicJourney}); (4) the participants gave feedback on design alternatives for interacting with the generated images in conversation.
  \label{fig:act}}
  \Description{This set of pictures shows the activities in two workshops. In sub-picture 1, three participants observe how to use the prototype, and the researchers manipulate the prototype equipment. In sub-picture 2, three participants brainstorm, and stickers mark key points on the table. In subfigure 3, a participant is experiencing a digital prototype with the help of a researcher. Subfigure 4 shows participants’ suggestions for different prototypes.}
\end{figure}

\textit{Group Design Discussion.} We then split those seven participants into two groups (Group 1: P1-4, Group 2: P5-7). Each facilitator conducted a semi-structured group interview to collect participants' feedback regarding their attitudes towards generative AI and the presented design features in the video prototype [Figure~\ref{fig:act} (1)], and to generate new design ideas through collaborative brainstorming [Figure~\ref{fig:act} (2)]. We closed the workshop by reflecting on and summarizing the design discussions with the participants.

\subsection{Workshop \romannum{2}: Examining Design Considerations}


\begin{figure*}[h]
  \centering
  \includegraphics[width=.95\linewidth]{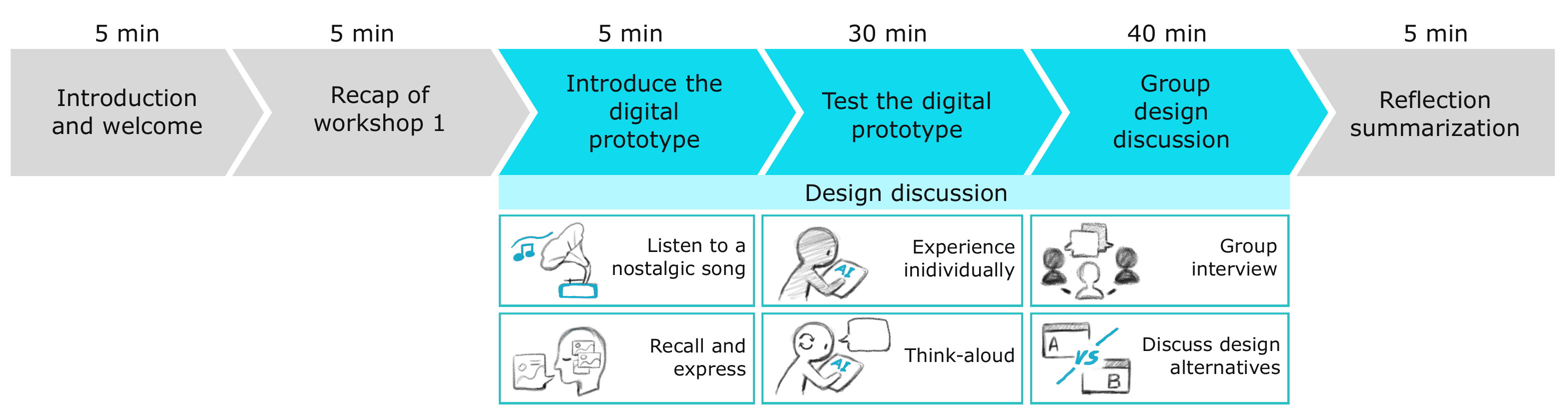}
  \caption{The overview of Workshop \romannum{2} including various activities, such as introducing and testing digital prototypes and group design discussion.}
  \label{fig:ws2}
  \Description{This flow chart describes the experimental process of the second workshop and includes various activities, such as introducing and testing digital prototypes and group design discussion.}
\end{figure*}

\begin{figure*}
  \centering
  \includegraphics[width=.90\linewidth]{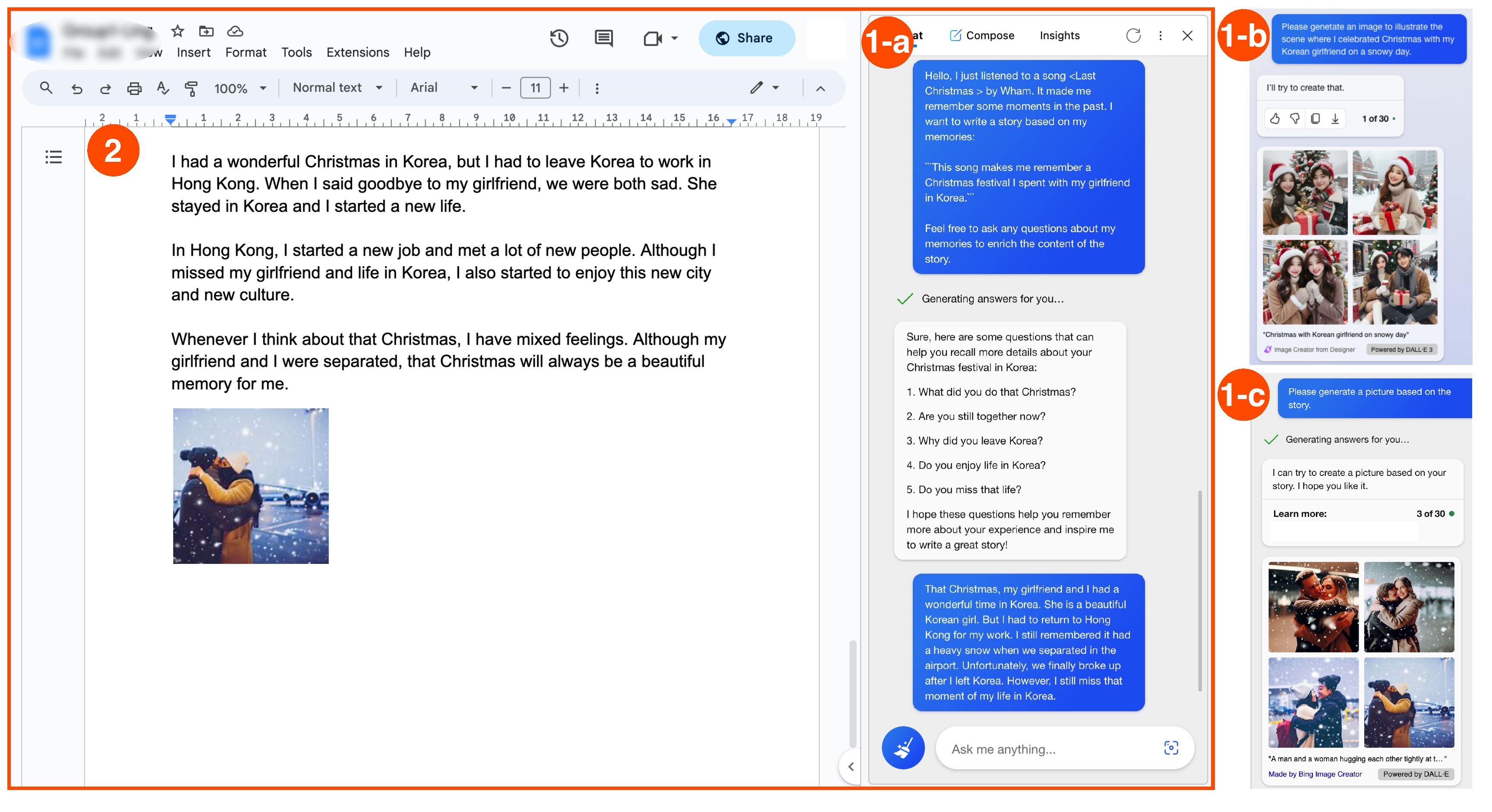}
  \caption{\wlrevised{The screenshots of the digital prototype (\textit{MusicJourney}). The user interface contains two parts: (1-a) the conversational interface with an example of generated questions to guide participants to think in music reminiscence, and (2) the text editor that presents a story generated based on the participant's conversation with Bing Chat. We also present two examples of generated images based on the recalled memory fragments (1-b) and based on the whole story (1-c). Note that the content displayed in the provided screenshots was translated from Chinese to English to facilitate comprehension.}}
  \label{fig:digital}
  \Description{The screenshots of the digital prototype (Music Journey). The user interface contains two parts: Subfigure 1-a shows the conversational agent with an example of generated questions to guide participants to think in music reminiscence. Subfigure 2 shows the text editor that presents a story generated based on participants’ conversations with Bing Chat. Subfigures 1-b and 1-c show the two examples of generating images, respectively, based on the recalled memory and based on the story.}
\end{figure*}

Based on the feedback in Workshop \romannum{1}, we refined our design concept and the setting of music reminiscence. For example, we decided to focus on the individual setting in light of participants' reluctance to share personal memories with strangers. Furthermore, given the design features discussed in Workshop \romannum{1}, we aimed to conduct another design workshop to examine the design considerations for a generative AI tool that could support music reminiscence by evaluating a digital prototype (\textit{MusicJourney}). The prototype was built on top of Bing Chat, a conversational agent powered by GPT-4. Compared with other LLM-based tools, Bing Chat can provide up-to-date information connected to the internet and generate creative images in conversation. These features allowed us to ask questions [Figure~\ref{fig:digital} (1-a)] and generate images [Figure~\ref{fig:digital} (1-b)] to help participants recall and reminisce about their life events linked to the played song. In addition, to better utilize Bing Chat to generate quality content \wlrevised{(e.g., questions to help participants recall their memories)}, we evaluated multiple prompts and found that a prompt about writing a story \wlrevised{based on reminiscence content could lead to the generation of }detailed and insightful questions. We provided a text editor that allows participants to view and edit the generated story about music-evoked memories [Figure~\ref{fig:digital} (2)].

Figure~\ref{fig:ws2} shows an overview of the Workshop \romannum{2}. The design discussions were centered on the evaluation of the digital prototype and design considerations for generative AI (e.g., interaction with the generative AI and user control for the generated content/images). Nine participants attended this workshop, six of whom participated in Workshop \romannum{1}. This workshop involved three facilitators and one collector. The workshop started with an introduction to the workshop and a recap of the activities carried out in Workshop \romannum{1}.

 
\textit{Introduce Digital Prototype.} We introduced the basic functions of the digital prototype and how users could interact with the prototype. We developed the digital prototype to support an activity where participants first listen to a nostalgia song and then recall and express their memories~\cite{istvandity2017combining}. \ycrevised{The digital prototype \textit{MusicJourney} operates as a web-based application accessed through a desktop browser. It comprises two primary components: Bing Chat, a conversational agent driven by large language models [Figure~\ref{fig:digital} (1-a)], and a text editor utilized to modify the generated story [Figure~\ref{fig:digital} (2)]. Initially, users are directed to commence searching for nostalgic music in Bing Chat. The music is then provided as a web link from YouTube, allowing users to play the music in the background. Meanwhile, users can share their feelings and memories with Bing Chat and respond to the questions generated by Bing Chat to recall more details [Figure~\ref{fig:digital} (1-a)]. Additionally, participants have the option to use Bing Chat to generate images based on the scenes that emerge in their memories [Figure~\ref{fig:digital} (1-b)]. Following Bing Chat's generation of a complete story, users can use the text editor to refine the story text and insert additional generated images [Figure~\ref{fig:digital} (1-c)] into the story.}

\textit{Test Digital Prototype.} We formed three groups (Group 1: P1, P2, P4; Group 2: P5-7; Group 3: P8-10) and asked each group member to test the digital prototype individually. Specifically, each participant listened to a nostalgic song they indicated in the pre-registration form. Afterward, the facilitator assisted the participant in typing our predefined prompt and their expressed memories in Bing Chat [Figure~\ref{fig:digital} (1-a)]. For example, the participant asked Bing Chat {``\textit{I just listened to a song. It made me remember some moments in the past. I want to write a story based on my memories of a Christmas festival I spent with my girlfriend in Korea. Feel free to ask any questions about my memories to enrich the story's content.}"}. Based on the generated questions, the participant can provide more details about the memories. For example, the participant answered some questions \textit{``Are you still together now?''} and \textit{``Why did you leave Korea?''}. Moreover, \wlrevised{each participant was guided to use Bing Chat to generate an image based on the recalled memories if they wanted to, which aimed to help them think about more details. To be more specific, the participant can ask Bing Chat first to generate a story based on the memory description and then generate an image based on the generated story or a sentence in the story.} Through a think-aloud protocol, we collected participants' experiences and opinions about using the prototype.

\textit{Group Design Discussion.} The design discussion was guided by a group semi-structured interview. We prepared the interview scripts focusing on the 
concrete and critical design considerations for generative AI in music reminiscence, some of which were informed by human-AI interaction guidelines ~\cite{amershi2019guidelines} and design principles for generative AI~\cite{weisz2023toward}. We made a booklet showing the design considerations and alternatives for conversation design (e.g., AI guidance, interaction modality, and explanations) and content generation (e.g., in respect of user control, required user input, and presentation). Facilitators collected participants' feedback for each of these design considerations and evaluated design alternatives against that of the digital prototype. For example, Figure~\ref{fig:act} (4) shows participants' feedback on the design of adjusting the generated image. In this example, participants can adjust the generated image using buttons providing feedback options (left), choosing one from four candidates (middle), or through conversation (right). The workshop concluded with a collective reflection and summary of the design discussions with the participants.

\subsection{Data Collection and Analysis}

The interviews in the two design workshops were audio-recorded and transcribed. We analyzed the transcripts with ATLAS.ti\footnote{https://atlasti.com/}, a tool for qualitative data analysis, using a reflexive thematic analysis method~\cite{braun2006using}. 
Specifically, we conducted our thematic analysis through the following four steps:
1) After each design workshop, researchers who attended the in-person workshops made detailed notes on the key ideas gathered from participants during the interviews. 
2) Two researchers independently went through the notes and coded them using an open coding technique, and then met to discuss and inductively produce the initial codes. 
3) Then, one researcher coded all the interview transcripts with the initial codes to ensure that the initial codes could capture all the interview data. The two researchers discussed any modifications of codes until they reached a consensus. 
4) After the initial coding, two researchers met to merge and group individual codes into categories, gain insights by building connections between the codes, and then develop themes related to our research questions. We performed the qualitative analysis in Chinese. Finally, we translated quotations from the interviews and reported our results in English.

\subsection{Positionality Statements}

In this research, our interdisciplinary team brings together expertise from the fields of human-computer interaction, artificial intelligence, and psychology research. We believe that generative AI technologies have the potential to support reminiscence among older adults and explore this direction from a human-computer interaction perspective. Four authors were involved in the data collection and analysis. Among them, three authors representing diverse genders worked together to perform the reflexive thematic analysis of the collected data, enabling us to explore different interpretations and perspectives. They all have experience co-designing technology with older adults in Hong Kong. Notably, one lead researcher has established a stable relationship with the community center since he collaborated with the center to run a design project about accessibility for older adults in 2021. These experiences have provided our research team with first-hand insights into the lives of older adults.


\subsection{Ethical Considerations}
 
The vulnerability of older adults led us to carefully consider the ethics of our study. We validated the schedule and outline of the design workshop with a social worker responsible for organizing daily activities for the members of this community center. The social worker helped us distribute an information sheet about the study to the participants one week before each workshop. On the day of each workshop, we also acquired written permission from participants for audio recording and photo taking. Additionally, we selected songs in our study by screening the music and associated memories they provided in the pre-registration form to try to avoid negative memories evoked by music. The social worker was also present \wlrevised{at the workshops to ensure that no participants were feeling distressed.} All facets of this research design, including voice recording, photography, and employment of generative AI, have been guided by ethical considerations. The Research Ethics Board of the first author's university approved the design protocol. \wlrevised{Participants could withdraw at any time, but no participants opted to withdraw from this study.}

\section{In-Depth Understanding of Older Adults}

We employed a reflexive thematic analysis approach to analyze the transcripts from Workshop \romannum{1}, aiming to gain insight into the perspectives of older adults regarding generated conversations and images for music reminiscence. The findings of our analysis provide a deep understanding of participants' subjective experiences, their acceptance of the initial design concepts presented in the video prototype, and their envisioned potential for generative AI in the context of music reminiscence.

\subsection{Perspectives on Generated Conversations}

\subsubsection{The tone of AI voice and the understanding of human emotions influence older adults' intention to communicate with AI} Although we used state-of-the-art technology to generate Cantonese speech in the video prototype \textit{AI-DJ}, some participants were not satisfied with the tone of the conversation and unwilling to communicate with such an AI system. For instance, P4 stated:

\begin{quote}
\textit{``I feel that the tone of the AI voice is a bit stiff, and it feels like the AI pronounces each word rather than communicating with emotion. Maybe the current technology can't do it yet.''} (P4)
\end{quote}

P2 agreed with this point and thought the tone of the AI voice should match the emotions conveyed by the conversation and music.

\begin{quote}
\textit{``When expressing unhappiness, I felt that the tone did not correspond to the emotion. I think the tone of AI voice and expression were not suitable for telling life stories. The tone was a bit stiff and didn't sound amiable.''} (P2)
\end{quote}

Additionally, another member of Group 1 mentioned that he was more willing to communicate with the AI if he felt that the AI could help him find more interesting songs and understand his feelings about these songs.

\begin{quote}
\textit{``If [AI] really understands me and provides me with a lot of good music to choose from, I will be very happy to spend time talking to [AI]. It could be my social time with AI. I usually have to think about what to do in the evening and which friends to find.''} (P3)
\end{quote}


\subsubsection{The generated conversations should reflect older adults' feelings and memories} Our objective is to assist participants in music reminiscence by offering additional information about the music, such as the story behind it and other listeners' comments regarding their emotions and memories~\cite{jin2023understanding}. However, presenting this information does not effectively support participants in their music reminiscence endeavors, as individuals in a group may not necessarily share the same feelings or memories associated with a particular song. For example, P2 and P4 noted:

\begin{quote}
\textit{``After a song is played, some background of the song is introduced. In fact, we already know the background very well, so basically, I don't need AI to tell me about this.''} (P2)
\end{quote}

\begin{quote}
\textit{``Because our main purpose is to listen to the music, I don't want to listen to other listeners' comments, especially if they are irrelevant to my memories. This is my personal listening experience, which may differ from others' experiences that I do not care about.''} (P4)
\end{quote}

Some participants also thought it was difficult for generative AI to moderate the conversations discussing divergent feelings and memories, especially in the group setting. For instance, P3 said:


\begin{quote}
\textit{``But my question is, what if three people are listening to a song, none of them know each other, and they all have different opinions? How could AI support all of us?''} (P3)
\end{quote}

\subsubsection{Individual conversations with AI are preferred over group conversation for music reminiscence} 

Participants expected not to be interrupted by irrelevant conversations while listening to music. For music reminiscence, they prefer chatting with AI alone rather than involving other unknown listeners because the individual conversation could spark deeper thoughts. For instance, P3 noted:

\begin{quote}
\textit{``In the individual setting, I can ask AI many questions, and he will immediately respond to my questions. If I am in a group, it will be difficult. I should have entirely different questions than another person. AI needs to deal with different questions/opinions and cannot go into depth.''} (P3)
\end{quote}

We considered the group setting for music reminiscence as it has been proven effective for mitigating common mental health issues for older adults~\cite{elias2015effectiveness,wang2007group}. However, in our demonstrated group setting, the \textit{AI-DJ} takes on the role of moderating the activity, which is distinct from the typical setting of real group reminiscence therapy where a social worker or therapist usually fulfills this role. The participants had several concerns about sharing their memories with other listeners, especially with strangers, including a lack of emotional resonance, online safety, and interruptions to the music listening. For example, P3 and P4 expressed difficulty in building emotional resonance:

\begin{quote}
\textit{``Even if people may like the same music, having a sense of resonance is important for group conversation. For example, those who like songs from the 1980s may have a common life experience to share with. Emotional resonance allows different people to share it. Emotional resonance means that you like the same song and have similar feelings about it.''} (P4)
\end{quote}


As older adults are vulnerable to cybersecurity attacks ~\cite{morrison2021older}, P1 also worried that she would be cheated by some people in a group conversation. 

\begin{quote}
\textit{``Worried that there are many scammers online, so I dare not to share too many personal things with them.''} (P1)
\end{quote}

Additionally, participants might engage more with other listeners' comments in group conversations than in individual conversations with AI, which could interrupt music listening. For example, P2 said:  
\begin{quote}
\textit{``I don't like AI talking too much while it plays music, which interrupts my music listening experience.''} (P2)
\end{quote}

Some participants were even reluctant to share their memories with their relatives because they have not often listened to music together, let alone discuss feelings and memories. For instance, P5 mentioned:

\begin{quote}
\textit{``It is difficult to explain my feelings clearly to others. I don't want to communicate with others ... I often listen to music alone. I even do not share my feelings and memories with my husband.''} (P5)
\end{quote}

P6 agreed and said:

\begin{quote}
\textit{``...my husband is dull and cannot understand my feelings about music. I don't like discussing music with him.''} (P6)
\end{quote}



\subsection{Perspectives on Generated Images}

\subsubsection{Feedback from all participants emphasized the importance of image quality to the reminiscence facilitated by these images.} Participants often evaluated the quality of the generated images by assessing the association between the depicted scenes and their own memories. For example, P2 said:

\begin{quote}
\textit{``If the picture has nothing to do with the scene in the past, I think it does not support me well in reminiscence.''} (P2)
\end{quote}

The image could be generated from a variety of sources, potentially impacting the support for reminiscence perceived by the participant. The video prototype utilizes diverse sources, including lyrics, the story behind the music, and personal memories to generate images. Some participants favored images generated based on the story behind the music because these images could strengthen the link between the music and their memories. For example, P2 mentioned:

\begin{quote}
\textit{``While playing the song ``The Private Eyes'', if the image shows the relevant scenes about the movie related to the song, I think it could help me recall some moments better.''} (P2)
\end{quote}

However, some participants thought generating quality images to match their described memories could be challenging. For example, P1 noted:

\begin{quote}
\textit{``Currently, I feel it is difficult for AI to generate the picture that matches my memory because AI cannot accurately visualize what I thought unless I tell it all the details.''} (P1)
\end{quote}

\subsubsection{People may ignore images while listening to music in some contexts.}
Moreover, because some participants may do daily activities, such as cooking, reading, and exercising while listening to music, they thought they might ignore the generated images if they were not asked to pay attention to them. For instance, P4 noted:

\begin{quote}
\textit{``Sometimes when I listen to music on YouTube, I won't even watch the video. Let alone view images when I listen to music while cooking. Rarely does everyone look at pictures simultaneously while listening to music unless you ask me to do so, right?''} (P4)
\end{quote}

\subsection{Concerns about Generative AI}
Although participants in Workshop~\romannum{1} did not experience generative AI for music reminiscence, the video prototype still raised their concerns about privacy, controllability, and the ability to understand humans. 


Similar to listening to music radio, users of the video prototype were expected to adhere to the conversational guidance provided by \textit{AI-DJ} to recall their memories triggered by the music. Nonetheless, certain participants expressed their desire to control \textit{AI-DJ} if the delivered content (e.g., music or conversation) fails to resonate with their emotions or feelings. For example, P2 noted:

\begin{quote}
\textit{``If my emotions are different from the emotions conveyed by the music, I would like to press a button, e.g., ``I am not interested in this song'', at the beginning of playing this song.''} (P2)
\end{quote}

\begin{quote}
\textit{``If my thoughts and feelings differ from what AI says, I may not be interested in listening to them anymore. So it's better if I have a choice to skip what is being said.''} (P2)
\end{quote}

The privacy concerns were centered on the uncertainty regarding the individuals authorized to access and read their personal data, as well as the inadvertent data leakage caused by unintentional actions. For example, P1 and P6 mentioned:

\begin{quote}
\textit{``Yes, I am worried because I don't know how AI will process my data. [For example,] what kind of people will hear my message?''} (P1)
\end{quote}


\begin{quote}
\textit{``I am worried that with one click of a button, it will reveal something I do not want others to know.''} (P6)
\end{quote}

In addition, P7 doubted AI's ability to understand their nuanced emotions and complicated stories of memories: 
\begin{quote}
\textit{``I wonder if AI is able to understand my nuanced emotions and complicated stories in my memory. If not, I would not like to share them with AI.''} (P7)
\end{quote}

\section{Design Considerations of Generative AI for Supporting Music Reminiscence}

In Workshop \romannum{2}, we conducted individual evaluations of the designed digital prototype \textit{MusicJourney} with nine older adults and then examined the design considerations for using generative AI to support music reminiscence through three group interviews. In the interviews with our participants, they generally felt that such a generative AI system prototype, which generated conversations and images based on their memories, could facilitate their reminiscence. For instance, P8 mentioned that the questions generated by AI can encourage them to reflect on their experiences from different perspectives during the reminiscing process:

\begin{quote}
\textit{``The [AI] system proactively asks me questions, which can guide my thinking and prompt me to consider different aspects.''} (P8)
\end{quote}

\wlrevised{The older adults in our study} also shared their detailed preferences and thoughts based on the experienced digital prototype and the design alternatives shown to them, which can inform the future design of generative AI tools for supporting music reminiscence. In the following, we will show the major themes that were identified through our thematic analysis of the three group interviews in Workshop \romannum{2}.

\subsection{Generative AI: AI Initiative and User Control}
\label{results-2-AI-Initiative}

\subsubsection{The guidance provided by generative AI enhances the reminiscing experience, yet older adults desire control over the direction of their reminiscences}
In our designed digital prototype, the system can generate questions to facilitate \wlrevised{older adults' reminiscence. When asked about our participants' opinions about the guidance offered by AI, they perceived that, with such guidance, their reminiscing process could be more pleasurable.} Also, conversational interaction with AI that mimics human conversation can increase their engagement as it may give older adults a feeling of communication with others, as noted by P2:  

\begin{quote}
\textit{``Reminiscence alone will make me feel bored and lonely ... When I saw other people experiencing this system, it felt like a conversation, and the questions that the AI asked them were not long. I felt the whole process would be quite comfortable. Then, when I tried it, I immediately fell in love with this AI system.}'' (P2)
\end{quote}

Although the support provided by AI was appreciated, our participants also emphasized the significance of retaining autonomy in determining the direction and depth of their reminiscence of past memories. For example, if AI inquires about a particular memory to facilitate reminiscence, older adults should be able to decide whether to reminisce in that direction according to their own wishes, as mentioned by P10:

\begin{quote}
\textit{``When the AI prompts me to share more, if I really want to add something, I will continue to talk. If I don't want to add anything, I will let the AI change questions.''} (P10)
\end{quote}

\subsubsection{AI systems are anticipated to seek input from older adults regarding the timing and content of generation.} 

Building upon the preceding discussion in terms of user control, participants were further prompted to discuss their expectations of the interaction with AI, particularly focusing on \textit{when} AI is expected to generate and \textit{what} AI could generate because it is important to determine the suitable timing and content based on user preferences when designing human-AI interaction~\cite{amershi2019guidelines}. In this regard, \wlrevised{some older adults} expressed that the AI system could ask about their opinions and seek their input (e.g., expected image content) before generation. For example, P2 told us:  

\begin{quote} 
\textit{``I hope the AI is more like a consultant. It can ask me if I need to generate an image using my story. If I say yes, he will generate it. If I say no, he will not generate it.''} (P2)
\end{quote}

Another participant, P4, desired more control during the interaction and expected that AI would generate images related to himself when he explicitly requested it:

\begin{quote}
\textit{``If AI wants to generate images related to me, I hope that I tell it when I need to generate images, instead of automatically generating them for me.''} (P4)
\end{quote}

Regarding the generated images, our participants indicated their willingness to provide feedback to the AI system. Then, the system could make adjustments accordingly. In terms of how to provide feedback on the generated images, during the interviews, we presented three design alternatives to the participants to elicit their preferences: (1) utilizing feedback buttons that include suggestions on how to adjust (e.g., ``including real persons''), (2) selecting from multiple generated images, and (3) expressing preferences using natural language. Most of the participants favored the first and the third options because they felt these two designs allowed them to provide more nuanced feedback. For example, P6 shared that: 

\begin{quote}
\textit{``After the AI automatically generates related images, we can ask it to make some modifications, such as deleting the moon or sun.''} (P6)
\end{quote}

In short, our participants generally expected a high level of control over interaction when using generative AI for music reminiscence, which can be achieved through questioning and feedback mechanisms. In the following, we will investigate our participants' more detailed thoughts on the design of generated conversation (Sections~\ref{results-2-Conversation-Input} and \ref{results-2-Conversation-Output}) and generated images (Section~\ref{results-2-ImageGen}) for supporting music reminiscence.

\subsection{Generated Conversation: User Input}
\label{results-2-Conversation-Input}

\subsubsection{Older adults prefer starting with simple descriptions and incrementally engaging in reminiscence about their memories through conversation with AI} 

Social context may influence older adults' behavior during reminiscence~\cite{lamme1993including}. For instance, sharing memories with a group of strangers and with family members would lead to different behavior than reminiscing alone; some may be willing to share a lot, while others may be more conservative. 
In our interview, we asked our participants how they would prefer to talk to the AI system about their memories, specifically about whether they preferred using extensive or brief descriptions or even keywords. Some participants said they would prefer to use short descriptions to share memories with the AI system and engage in conversation with AI to recall more details. For example, P8 said:

\begin{quote}
\textit{``I hope that the input of my content can be just simple key points at the beginning. After I finish talking about some key points, AI can prompt me to provide more details ...''} (P8)
\end{quote}



Moreover, participants discussed that the content of reminiscence may be related to the emotions evoked by the listened music or depend on the mood of that day. For example, P1 said:
\begin{quote}
\textit{``I feel that the amount of content I recall is related to my mood. If I am happy that day, I may talk more. If I happen to be tired one day, I may talk less."} (P1)
\end{quote}

Following this conversation, P4, who belonged to the same group as P1, suggested that AI should adapt to diverse needs by accommodating inputs with varying degrees of detail: 
 
\begin{quote}
\textit{``I think AI should accept content input with varying levels of detail as different users' needs are different."} (P4)
\end{quote}


\subsubsection{Regarding interaction modality, multi-modal input is preferable in the context of music reminiscence.}
When asked about participants' preferences for interaction modality (e.g., either using speech or typing), our participants thought that voice interaction is more convenient, but there will be some issues in the context of music reminiscence. For example, P4 mentioned that using voice interaction when listening to music may cause interruptions: 

\begin{quote}
\textit{``I'm worried that the conversation between the AI and me will interrupt me listening to music, which is the main problem right now.''} (P4)
\end{quote}

In the same group, P2 expressed a degree of discomfort if she needed to verbally share her memories:

\begin{quote}
\textit{``I feel a little embarrassed expressing my feelings through voice.''} (P2)
\end{quote}

Compared to voice input, P2 thought that text input may sometimes allow her to express her complex feelings:

\begin{quote}
\textit{``I think words can sometimes express richer and deeper emotions.''} (P2)
\end{quote}

Considering that each individual may have their own preferences when sharing their memories, multi-modal input would be a better choice when designing the user interface to support music reminiscence.

\subsection{Generated Conversation: AI Output}
\label{results-2-Conversation-Output}

\subsubsection{Concerning AI-generated content, older adults favored concise conversations and expected the length of AI-generated content to be commensurate with their input.}
When discussing the expectations on the length of AI-generated conversation, some participants preferred shorter output, which can be short questions or concise descriptions, to stimulate their memories in music reminiscence because they may feel tired when reading long texts. This can be attributed to the fact that older people may be more prone to age-related eye problems or vision issues due to the natural aging process~\cite{horowitz2004prevalence}. For example, P4 said:

\begin{quote}
\textit{``I don't want to read too many words. As my eyes are presbyopic, I am tired from reading this.''} (P4)
\end{quote}

Moreover, participants in Group 1 engaged in an interesting discussion, revealing the importance of carefully considering the length of AI-generated content in the context of reminiscence support. P2 thought the length of AI-generated content should be commensurate with their input; otherwise, they would feel they were not being respected. 

\begin{quote}
\textit{``If the AI responds much shorter than my input, I definitely don't like it.''} (P2)
\end{quote}

P4 further added that he would expect AI to possess the ability to discern his requirements from the information he provided and to give pertinent responses: 
\begin{quote}
\textit{``I hope that AI will reply to me with detailed content after I describe it to him in detail. At the same time, when I describe simple content to him, it can also identify my needs from my description and give relevant responses.''} (P4)
\end{quote}

\subsubsection{AI-generated conversations are expected to be human-like, personalized, and authentic.} Similar to the feedback from Workshop \romannum{1},  participants of Workshop \romannum{2} expected that the generated conversation would exhibit a certain degree of human-like characteristics and personalization. In addition, P2 mentioned that the generated stories by AI should be authentic:

\begin{quote}
\textit{``I think AI's text output does not need to be very detailed, but its tone needs to be human-like so that I will feel happy when I hear it. I think the current level of humanization of AI is enough, which is pretty good ... When it comes to generating stories, I hope that the story generated by AI is related to my own story, a true story, rather than a story created by AI itself.''} (P2)
\end{quote}

One possible reason for this could be that authentic stories can resonate with their past life experiences, allowing them to connect emotionally with the content. Such emotional connection may facilitate the reminiscence process, making it more engaging and effective in evoking memories.

\subsubsection{Regarding whether AI should provide explanations for its generated content, older adults have varying preferences.}
As explanations have been shown to influence users' experience and perceptions of using AI in various tasks \cite{wang2021explanations}, we were also interested in seeing whether older adults may want explanations from AI regarding the content it generates, particularly when the AI system is designed to facilitate their reminiscence activities. In this regard, our participants hold different perspectives. Some thought it would not be necessary to provide such explanations because the explanations are not related to their reminiscing, as P1 said:  

\begin{quote}
\textit{``There is no need to tell me so much. I do not wish for any explanation, as I feel that it has no direct relationship with my own memories. I prefer not to know.''} (P1)
\end{quote}

However, P4 disagreed with P1, noting that if AI can provide explanations for the generated content in a manner that aligns with his preferences, it may enhance his perception of the generated content:

\begin{quote}
\textit{``AI can learn my preferences. If the AI interprets the generated content according to my preferences, I will feel that the generated content is more personalized.''} (P4)
\end{quote}

P8 also expressed a desire for AI explanations because she wanted to know the degree to which the AI understood her:

\begin{quote}
\textit{`` [With explanations,] I can know how well AI understands what I say.''} (P8)
\end{quote}

\subsection{Generated Images}
\label{results-2-ImageGen}

\subsubsection{Older adults anticipated that the generated images for music reminiscence could be more personally relevant to them and authentic.}
Regarding AI-generated images, in Workshop \romannum{1}, our participants indicated that the quality and relevance of the generated images may impact their willingness to use  AI for reminiscence, which was also revealed in Workshop \romannum{2} when they experienced the digital prototype to perform music reminiscence. Specifically, in the interviews, when questioned about the content (i.e., music background story, music lyrics, and personal stories) they would like to use to generate images to support their reminiscence, most of the participants preferred to use their own stories to generate images because they are personally connected to them, as said by P10:

\begin{quote}
\textit{``I think it would be better for AI to generate images based on my own stories because such images are more relevant to me.''} (P10)
\end{quote}

P4 expressed her wish that AI can generate images that include herself: 
\begin{quote}
\textit{``I want AI to generate images about myself. Just like when I want to recall something, I would look for photos on my phone that I rarely browse. The images generated by AI that include myself will make me feel better.''} (P4)
\end{quote}

In addition, as some participants encountered several generated images with foreign backgrounds or foreign people, they believe that the generated images should align with cultural background to better support their reminiscence experiences, as P2 noted:

\begin{quote}
\textit{``I think the content generated by AI needs to be differentiated between different regions. For example, if we are in Hong Kong, the content generated by AI needs to be related to our living background. Otherwise, the content generated by AI will be inconsistent with our reality.''} (P2)
\end{quote}

The results above indicate that our participants expected AI-generated content to possess authenticity and relevance for facilitating more effective stimulation of their memories.

\subsubsection{Older adults had privacy concerns, but some indicated their willingness to share data if they trust the AI}

While our participants appreciated the generated images for aiding their reminiscence, this was accompanied by privacy concerns regarding their personal data. Some participants were willing to share their data during the interview, depending upon their perception of the AI system's trustworthiness. For example, P4 said:

\begin{quote}
\textit{``I guess I won't tell AI everything about myself in the beginning, but if all my concerns [about privacy] are eliminated after some time, I will share more with AI.''} (P4)
\end{quote}

In line with broader AI research, when designing generative AI systems aimed at enhancing music-based reminiscence among older adults, researchers should take their privacy concerns into account throughout the design process.

\section{Discussion}
Music-based reminiscence has emerged as an effective method for promoting mental well-being among older adults~\cite{engelbrecht2023music,elias2015effectiveness}. 
However, older individuals may face challenges in engaging with music-based reminiscence due to age-related cognitive issues, including memory loss, difficulty concentrating, and susceptibility to distractions~\cite{lin2013hearing}. \wlrevised{Recognizing the tremendous potential of generative artificial intelligence (AI) in assisting individuals with complex and creative tasks~\cite{vert2023will,hughes2021generative}, we believe that generative AI can be a potentially supportive tool to enhance
older adults’ reminiscence experiences, e.g., by generating relevant content to trigger their memories. Thus, this study sets out to investigate \textit{whether} and \textit{how} generative AI can be designed to support reminiscence for older adults. Through interviews with social workers and two design workshops with older adults, we have explored how AI-generated content, such as questions in conversations and images, could help older adults recall more details of their past experiences and engage them in the reminiscing process. This research also provides new insights into older adults' attitudes toward using emerging technology (i.e., generative AI) for their reminiscing activities as well as their individual preferences, which could be considered in the future design of generative AI systems specifically tailored for music reminiscence among older adults. In this section, we present our understanding of the perspectives held by \ycrevised{the older adults in our study} towards generative AI, along with a set of design considerations aimed at leveraging generative AI to facilitate music reminiscence experiences for this older population. }

\subsection{Older Adults' Perspectives on Generative AI}

\textbf{Older adults acknowledge the practical assistance offered by AI and express a desire for more emotionally engaging and human-like interaction with generative AI}. \wlrevised{The older adults in our study believed that the guidance of AI, including generated questions and images, and human-like conversational interaction, can make the reminiscence process more pleasurable. One aspect highlighted by the older adults was the tone of voice of the designed AI; they felt it was important to match the tone with their feelings during music reminiscence. This may indicate that older adults would seek a deeper emotional connection and resonance with the generative AI system, reflecting the importance of humanness and empathy in communication between older adults and chatbots~\cite{rapp2021human}.  These findings suggest that incorporating emotional and human-like elements (e.g., self-introduction, adaptive voice tone and response speed, and echoing users' responses~\cite{rhim2022application}) into communication with generative AI might create a more engaging and satisfying music reminiscence for older adults. Despite the desire for human-like interactions, the participating older adults tended to perceive the generative AI system as a supportive tool rather than an intimate companion, which was also demonstrated in a companion chatbot for older adults~\cite{dosovitsky2021bonding}; they were inclined to use our conceptualized generative AI system for practical purposes of music reminiscence rather than seeking companionship. This finding aligns with previous studies on the role that technology often plays in the daily lives of older adults~\cite{rogers2003can}, which also showed that older adults typically use technology as pragmatic tools to enhance their daily activities, access information, or connect with others~\cite{selwyn2003older,garattini2012linking}.}

\textbf{Older adults prefer individual and deep conversations with AI when engaging in reminiscence activities triggered by music.} \wlrevised{To be specific, the older adults in our study tended to appreciate the guidance of the generative AI that supports them in reflecting on their life experiences triggered by the listened music, and to go deeper into their own memories through an \textit{introspective} and \textit{personal} reminiscence process~\cite{molinari1985life}. We also found that \wlrevised{some of the participating older adults} were reluctant to engage in group conversations with other listeners to share their experiences of music reminiscence. This reluctance might stem from privacy concerns regarding sharing with unfamiliar people or the effect of their different life experiences on impeding emotional resonance.} Some participants noted that they were more likely to accept a group setting where they could converse with trusted individuals, such as family members or close friends. For example, ``\textit{I think I would like to share my feelings and memories about music with my friends. So it would be great to invite my friends to join the music reminiscence activity.}'' (P1) \wlrevised{Therefore, in the future design of generative AI for music-based reminiscence among older adults, the interaction design may consider prioritizing a one-on-one interaction format and providing more personalized and empathetic responses~\cite{ma2021one,lin2020caire}, while as an alternative, such systems could also offer a controllable group setting where older adults can share their reminiscence experiences exclusively with trusted individuals. As the current study focuses on the design of generative AI for individual use based on most participants' preferences, it will be interesting to see how to best leverage generative AI in a group setting in future studies, e.g., in group reminiscence therapy~\cite{elias2015effectiveness}. For instance, generative AI might be used to facilitate therapists or social workers to engage older adults in the reminiscing process. 
Furthermore, regarding music reminiscence, some participating older adults expected AI to provide personalized music recommendations that can help them find music related to their past experiences, as they may sometimes forget the specific music they listened to in the past. As the current study primarily focuses on generative AI for music reminiscence, our exploration of music recommendation is limited. In future investigations, additional research is needed to better understand how to provide such support in music-based reminiscence AI systems~\cite{bermingham2013automatically}. For example, to further engage older adults in interacting with AI systems, future design may incorporate conversational recommendations and guidance to help older adults find their familiar songs~\cite{narducci2020investigation, cai2021critiquing} to recall their memories~\cite{chi2023listen}.}


\wlrevised{\textbf{Older adults have concerns about privacy risks and may have reservations about sharing personal memories or experiences with AI systems and unfamiliar individuals}. As reminiscence activities may trigger very personal stories, our participating older adults have privacy concerns about using the generative AI systems because they are not sure who will use the data and how their data will be used. Privacy concerns with emerging technologies among older adults have also been observed in other research in smart IoT devices~\cite{schroeder2022data}, mobile health applications \cite{ wang2019technology, matthew2016designing}, and socially assistive robots \cite{yusif2016older,pino2015we}, which may to some extent impact older adults' trust and intention to adopt these technologies in their daily lives. Given that some individual memories are highly personal and may contain sensitive personal information, it is crucial to take into account the privacy concerns in designing generative AI systems for older adults to perform music-based reminiscence. To ensure the protection of data privacy and build trust toward AI among older adults, future design should consider providing transparent and easy-to-understand privacy notices to older adults so that they could be aware of what data would be collected and used by the AI system and then make informed decisions~\cite{feng2021design}. It is also suggested to design accessible privacy controls for older adults by improving their findability and actionability~\cite{im2023less}. For example, the presentation of privacy choices (e.g.,  what options they have, and how to tell the system about their privacy decisions such as who to share and what data to share) should be easy to find and operate for older adults. Additionally, a previous study shows that people's privacy concerns and awareness are positively influenced by their digital literacy~\cite{park2013digital}. It is important to note that, in our study, the majority of our participants demonstrated good digital literacy, possibly making them more conscious of privacy issues than those with lower digital literacy when using generative AI for music reminiscence. }



\subsection{Generative AI for Music-Based Reminiscence}
Based on the two design workshops, we identified several critical aspects that warrant careful consideration when designing generative AI systems for music-based reminiscence support among older adults.

\textbf{Future design of AI systems for music-based reminiscence support among older adults should carefully balance AI initiative and user autonomy while actively seeking user input.} 
\wlrevised{Our work has provided some initial exploration of how the guidance provided by generative AI could stimulate memory recall, foster engaging conversations, and support detailed storytelling among the participating older adults}. While appreciating this reminiscence support, \wlrevised{the older adults in our study also expected a high level of control over the direction of their reminiscences}, which can be explained by the fact that reminiscence is closely related to individual characteristics (e.g., personality traits) and contextual factors (e.g., culture) \cite{westerhof2014celebrating}. Moreover, \wlrevised{the older adults expected} AI systems to actively seek their input, particularly regarding when to generate and what to generate, allowing them to reminisce at their own pace. They might experience a greater sense of empowerment by actively participating in shaping their reminiscence experience, which can contribute to their overall well-being~\cite{bohlmeijer2007effects}. In light of these findings, \wlrevised{we think AI reminiscence systems should engage in collaboration with older adults,} taking their preferences into account and aligning with their individual reminiscence directions. Such a human-AI collaboration approach may ensure that music reminiscence remains a \wlrevised{mixed-initiative (including both user-initiated and AI-initiated actions) and pleasurable activity for older adults,} with AI acting as a supportive facilitator. Therefore, in generative AI systems dedicated to supporting music reminiscence among older adults, it is essential to strike a thoughtful equilibrium between AI initiative and user autonomy. Additionally, these systems should proactively seek and incorporate user input from older adults, tailoring the experience to their needs.



\textbf{Supporting adaptive user interaction in generative AI systems is crucial for engaging older adults in music reminiscence and reducing the cognitive effort required to recall memories.} \ycrevised{The older adults in our study} exhibited a preference for performing reminiscence by briefly describing their memories during the interaction with AI and progressively delving deeper into their past memories through AI-guided conversations. Also, in music reminiscence, the emotion evoked by the listened music may influence their reminiscence, potentially prompting the recall of more or fewer memories~\cite{janata2007characterisation,istvandity2017combining}. These observations reveal the significance of adaptive user interaction, where AI systems should be adapted to the pace of older adults, dynamically adjusting their prompts and engagement levels,  e.g., based on the current emotional state of older adults. This adaptive approach may allow a gradual and engaging exploration of their memories in music reminiscence. Another observation from our study is that, regarding interaction modality, multi-modal input (e.g., voice and text inputs) that may satisfy different older adults' requirements is preferable when designing the AI system to support music reminiscence. Some \wlrevised{participating older adults preferred using voice input, especially in a private setting, as it may allow them to express their memories freely and naturally without disturbing others~\cite{stigall2019older}}. Conversely, \wlrevised{some older adults felt that text input was more suitable,} especially if they wanted to convey nuanced emotions or when they encountered difficulty with speech. Previous research also suggested that written expression may enable people to articulate their thoughts more precisely and more clearly~\cite{smyth1998written}, so encouraging text input may help older adults capture their memories with greater clarity in their reminiscing process.







\textbf{Tailoring generated content for music reminiscence among older adults requires concise and sincere conversations, personalized and unbiased content, and a balance between authenticity and creativity.}

\begin{itemize}
    \item \textit{Concise and sincere conversations}: \ycrevised{Some older adults in our study expressed a preference for concise rather than lengthy conversations} when engaging with AI-generated content. Concise conversations are easier to digest and may reduce cognitive fatigue in older adults~\cite{ryu2020simple}, which could allow them to focus on specific memories of the past and enhance their reminiscence experiences. This highlights the importance of conveying concise and to-the-point content and avoiding overly lengthy responses when designing the AI system to support older adults' reminiscence. \ycrevised{Nevertheless, AI responses are also expected to align with the input of users, reflecting a proportional level of detail to convey sincerity in conversations. Hence, it is recommended that AI adapt its responses based on the specific reminiscence content shared by older adults, demonstrating respect for their engagement.}

    \item \textit{Personalized content}: \wlrevised{The participating older adults expressed} a desire for more personalized content relevant to their own life experiences. Some \wlrevised{older adults} also thought they would be more engaged and motivated to reminisce about their past if AI could generate images based on their personal data, such as portrait images and love-life stories, while others have privacy concerns about this. This suggests that incorporating richer user-provided information can help generate content that resonates with their memories, making the reminiscence experience more meaningful and effective. Like other user demographics, older adults have concerns regarding the privacy and security of their personal data~\cite{hoofnagle2010different}. \wlrevised{As discussed in the previous section,} when designing AI systems to support the reminiscence of older adults, ensuring robust privacy protection should also be considered in the design and development phase. 
    
    \item \textit{Unbiased content}: This study also pointed out the importance of avoiding biases in the generated content, which was also discussed in the field of generative art~\cite{srinivasan2021biases}. \wlrevised{The participating older adults} mentioned that it is essential for the AI system to present images and other content that accurately represent their race and background. Stereotypes and biases in the generated content can significantly impact their reminiscence experiences. In the future design of generative AI to support reminiscence, it is crucial to develop robust mechanisms that prevent biased content generation, \wlrevised{ensuring a more inclusive and respectful experience for older adults.}

    \item \textit{Authenticity vs. Creativity}: \wlrevised{Some older adults in our study expected} AI-generated conversations to be authentic and closely aligned with their actual memories. However, some \ycrevised{older adults} were interested in seeing moderate creative elements generated by AI based on their memories. Future design may consider striking a balance between authenticity and creativity, which might enhance \wlrevised{older adults'} engagement in the reminiscing process. Providing user control or customization options that allow them to adjust the level of creativity of the generative AI system could be a valuable feature.
\end{itemize}

\section{Limitations \& Future Work}

This work has four limitations.
\textit{First}, given the difficulties in co-designing with older adults~\cite{sakaguchi2021co}, we facilitated some activities and discussions in our workshops. For example, in our design Workshop \romannum{2}, facilitators assisted older adults in using generative AI, e.g., rephrasing and typing their inputs, which may obscure some of the difficulties that older adults may actually encounter when they interact with the digital prototype~\cite{chu2022digital}. To some extent, these efforts may restrict the participation of older adults in the design process.
\textit{Second}, we conducted our study with older adults living in Hong Kong, China. Therefore, our findings may be influenced by the geographic and cultural context of that specific region. It would be valuable to conduct similar studies with older adults in different countries/regions to validate the generalizability of our findings. \ycrevised{\textit{Third}, the majority of our participants were well-educated and possessed a certain level of digital literacy. Given that the level of digital literacy may impact the concerns and preferences of older adults when they use AI technologies~\cite{chattaraman2019should}, the older adults in our study may have a better understanding of generative AI and a greater ability to engage with it.} \textit{Fourth}, our design study was conducted with a relatively small sample size.

For future work, we plan to develop an application based on our study findings and deploy it in the wild. Specifically, we aim to conduct a longitudinal study involving a larger sample size (e.g., with about 50 older adults) to validate and refine our findings in a real-world setting~\cite{chamberlain2012research}. Moreover, we foresee the potential of generative AI for reminiscence triggered by other prompts, such as photos, household items, and personal recordings~\cite{o1998reminiscence}. Therefore, we intend to investigate how well our findings could be generalized to other reminiscence activities based on different triggers.


\section{Conclusions}
With the advent of large language models, a growing number of generative AI applications have been developed to enhance human capabilities in various tasks. Nonetheless, limited attention has been paid to exploring the potential impact of generative AI on older adults. This study aims to understand older adults' perspectives on generative AI for a beneficial activity for their mental well-being -  music-based reminiscence. We conducted two design workshops with older adults to respectively evaluate two design prototypes (a video prototype and a digital prototype) of generative AI for music reminiscence. The participants of our workshops viewed generative AI as a powerful assistant as it could help them recall more details and diverse memories in music reminiscence. However, they desired greater autonomy, personalization, and authenticity in the AI-generated content. Additionally, we analyzed our findings from the perspective of human-computer interaction (HCI) practitioners and derived design considerations that can inform the design of generative AI applications to support music reminiscence among older adults. We hope this study will inspire further design research in the realm of older adults' health and well-being.

\begin{acks}
  This work was supported by China Hong Kong Research Grants Council (RGC) GRF project (RGC/HKBU12201620), and Hong Kong Baptist University IG-FNRA project (RC-FNRA-IG/21-22/SCI/01) and Start-up Grant (RC-STARTUP/21-22/23). This work was also supported, in part, by Science Foundation Ireland grant 13/RC/2094\_P2.
\end{acks}

\bibliographystyle{ACM-Reference-Format}
\bibliography{refs}


\begin{thebibliography}{94}


\ifx \showCODEN    \undefined \def \showCODEN     #1{\unskip}     \fi
\ifx \showDOI      \undefined \def \showDOI       #1{#1}\fi
\ifx \showISBNx    \undefined \def \showISBNx     #1{\unskip}     \fi
\ifx \showISBNxiii \undefined \def \showISBNxiii  #1{\unskip}     \fi
\ifx \showISSN     \undefined \def \showISSN      #1{\unskip}     \fi
\ifx \showLCCN     \undefined \def \showLCCN      #1{\unskip}     \fi
\ifx \shownote     \undefined \def \shownote      #1{#1}          \fi
\ifx \showarticletitle \undefined \def \showarticletitle #1{#1}   \fi
\ifx \showURL      \undefined \def \showURL       {\relax}        \fi
\providecommand\bibfield[2]{#2}
\providecommand\bibinfo[2]{#2}
\providecommand\natexlab[1]{#1}
\providecommand\showeprint[2][]{arXiv:#2}

\bibitem[Amershi et~al\mbox{.}(2019)]%
        {amershi2019guidelines}
\bibfield{author}{\bibinfo{person}{Saleema Amershi}, \bibinfo{person}{Dan Weld}, \bibinfo{person}{Mihaela Vorvoreanu}, \bibinfo{person}{Adam Fourney}, \bibinfo{person}{Besmira Nushi}, \bibinfo{person}{Penny Collisson}, \bibinfo{person}{Jina Suh}, \bibinfo{person}{Shamsi Iqbal}, \bibinfo{person}{Paul~N Bennett}, \bibinfo{person}{Kori Inkpen}, {et~al\mbox{.}}} \bibinfo{year}{2019}\natexlab{}.
\newblock \showarticletitle{Guidelines for human-AI interaction}. In \bibinfo{booktitle}{\emph{Proceedings of the 2019 chi conference on human factors in computing systems}}. \bibinfo{pages}{1--13}.
\newblock


\bibitem[Axtell and Munteanu(2019)]%
        {axtell2019photoflow}
\bibfield{author}{\bibinfo{person}{Benett Axtell} {and} \bibinfo{person}{Cosmin Munteanu}.} \bibinfo{year}{2019}\natexlab{}.
\newblock \showarticletitle{PhotoFlow in action: picture-mediated reminiscence supporting family socio-connectivity}. In \bibinfo{booktitle}{\emph{Extended Abstracts of the 2019 CHI Conference on Human Factors in Computing Systems}}. \bibinfo{pages}{1--4}.
\newblock


\bibitem[Baidoo-Anu and Owusu~Ansah(2023)]%
        {baidoo2023education}
\bibfield{author}{\bibinfo{person}{David Baidoo-Anu} {and} \bibinfo{person}{Leticia Owusu~Ansah}.} \bibinfo{year}{2023}\natexlab{}.
\newblock \showarticletitle{Education in the era of generative artificial intelligence (AI): Understanding the potential benefits of ChatGPT in promoting teaching and learning}.
\newblock \bibinfo{journal}{\emph{Available at SSRN 4337484}} (\bibinfo{year}{2023}).
\newblock


\bibitem[Bejan et~al\mbox{.}(2018)]%
        {bejan2018virtual}
\bibfield{author}{\bibinfo{person}{Alexander Bejan}, \bibinfo{person}{Markus Wieland}, \bibinfo{person}{Patrizia Murko}, {and} \bibinfo{person}{Christophe Kunze}.} \bibinfo{year}{2018}\natexlab{}.
\newblock \showarticletitle{A virtual environment gesture interaction system for people with dementia}. In \bibinfo{booktitle}{\emph{Proceedings of the 2018 ACM Conference Companion Publication on Designing Interactive Systems}}. \bibinfo{pages}{225--230}.
\newblock


\bibitem[Bermingham et~al\mbox{.}(2013)]%
        {bermingham2013automatically}
\bibfield{author}{\bibinfo{person}{Adam Bermingham}, \bibinfo{person}{Julia O'Rourke}, \bibinfo{person}{Cathal Gurrin}, \bibinfo{person}{Ronan Collins}, \bibinfo{person}{Kate Irving}, {and} \bibinfo{person}{Alan~F Smeaton}.} \bibinfo{year}{2013}\natexlab{}.
\newblock \showarticletitle{Automatically recommending multimedia content for use in group reminiscence therap}. In \bibinfo{booktitle}{\emph{Proceedings of the 1st ACM international workshop on Multimedia indexing and information retrieval for healthcare}}. \bibinfo{pages}{49--58}.
\newblock


\bibitem[Bluck and Levine(1998)]%
        {bluck1998reminiscence}
\bibfield{author}{\bibinfo{person}{Susan Bluck} {and} \bibinfo{person}{Linda~J Levine}.} \bibinfo{year}{1998}\natexlab{}.
\newblock \showarticletitle{Reminiscence as autobiographical memory: A catalyst for reminiscence theory development}.
\newblock \bibinfo{journal}{\emph{Ageing \& Society}} \bibinfo{volume}{18}, \bibinfo{number}{2} (\bibinfo{year}{1998}), \bibinfo{pages}{185--208}.
\newblock


\bibitem[Bohlmeijer et~al\mbox{.}(2007)]%
        {bohlmeijer2007effects}
\bibfield{author}{\bibinfo{person}{Ernst Bohlmeijer}, \bibinfo{person}{Marte Roemer}, \bibinfo{person}{Pim Cuijpers}, {and} \bibinfo{person}{Filip Smit}.} \bibinfo{year}{2007}\natexlab{}.
\newblock \showarticletitle{The effects of reminiscence on psychological well-being in older adults: A meta-analysis}.
\newblock \bibinfo{journal}{\emph{Aging and Mental Health}} \bibinfo{volume}{11}, \bibinfo{number}{3} (\bibinfo{year}{2007}), \bibinfo{pages}{291--300}.
\newblock


\bibitem[Braun and Clarke(2006)]%
        {braun2006using}
\bibfield{author}{\bibinfo{person}{Virginia Braun} {and} \bibinfo{person}{Victoria Clarke}.} \bibinfo{year}{2006}\natexlab{}.
\newblock \showarticletitle{Using thematic analysis in psychology}.
\newblock \bibinfo{journal}{\emph{Qualitative research in psychology}} \bibinfo{volume}{3}, \bibinfo{number}{2} (\bibinfo{year}{2006}), \bibinfo{pages}{77--101}.
\newblock


\bibitem[Bryant et~al\mbox{.}(2005)]%
        {bryant2005using}
\bibfield{author}{\bibinfo{person}{Fred~B Bryant}, \bibinfo{person}{Colette~M Smart}, {and} \bibinfo{person}{Scott~P King}.} \bibinfo{year}{2005}\natexlab{}.
\newblock \showarticletitle{Using the past to enhance the present: Boosting happiness through positive reminiscence}.
\newblock \bibinfo{journal}{\emph{Journal of Happiness Studies}}  \bibinfo{volume}{6} (\bibinfo{year}{2005}), \bibinfo{pages}{227--260}.
\newblock


\bibitem[Butler(1963)]%
        {butler1963life}
\bibfield{author}{\bibinfo{person}{Robert~N Butler}.} \bibinfo{year}{1963}\natexlab{}.
\newblock \showarticletitle{The life review: An interpretation of reminiscence in the aged}.
\newblock \bibinfo{journal}{\emph{Psychiatry}} \bibinfo{volume}{26}, \bibinfo{number}{1} (\bibinfo{year}{1963}), \bibinfo{pages}{65--76}.
\newblock


\bibitem[Cai et~al\mbox{.}(2021)]%
        {cai2021critiquing}
\bibfield{author}{\bibinfo{person}{Wanling Cai}, \bibinfo{person}{Yucheng Jin}, {and} \bibinfo{person}{Li Chen}.} \bibinfo{year}{2021}\natexlab{}.
\newblock \showarticletitle{Critiquing for music exploration in conversational recommender systems}. In \bibinfo{booktitle}{\emph{26th International Conference on Intelligent User Interfaces}}. \bibinfo{pages}{480--490}.
\newblock


\bibitem[Cai et~al\mbox{.}(2023)]%
        {chi2023listen}
\bibfield{author}{\bibinfo{person}{Wanling Cai}, \bibinfo{person}{Yucheng Jin}, \bibinfo{person}{Xianglin Zhao}, {and} \bibinfo{person}{Li Chen}.} \bibinfo{year}{2023}\natexlab{}.
\newblock \showarticletitle{“Listen to Music, Listen to Yourself”: Design of a Conversational Agent to Support Self-Awareness While Listening to Music}. In \bibinfo{booktitle}{\emph{Proceedings of the 2023 CHI Conference on Human Factors in Computing Systems}}. \bibinfo{pages}{1--19}.
\newblock


\bibitem[Cappeliez et~al\mbox{.}(2008)]%
        {cappeliez2008functions}
\bibfield{author}{\bibinfo{person}{Philippe Cappeliez}, \bibinfo{person}{Marilyn Guindon}, {and} \bibinfo{person}{Annie Robitaille}.} \bibinfo{year}{2008}\natexlab{}.
\newblock \showarticletitle{Functions of reminiscence and emotional regulation among older adults}.
\newblock \bibinfo{journal}{\emph{Journal of Aging Studies}} \bibinfo{volume}{22}, \bibinfo{number}{3} (\bibinfo{year}{2008}), \bibinfo{pages}{266--272}.
\newblock


\bibitem[Caprani et~al\mbox{.}(2005)]%
        {caprani2005remember}
\bibfield{author}{\bibinfo{person}{Niamh Caprani}, \bibinfo{person}{Nuala Dwyer}, \bibinfo{person}{Kim Harrison}, {and} \bibinfo{person}{Karen O'Brien}.} \bibinfo{year}{2005}\natexlab{}.
\newblock \showarticletitle{Remember when: development of an interactive reminiscence device}. In \bibinfo{booktitle}{\emph{CHI'05 Extended Abstracts on Human Factors in Computing Systems}}. \bibinfo{pages}{2070--2073}.
\newblock


\bibitem[Car{\'o}s et~al\mbox{.}(2020)]%
        {caros2020automatic}
\bibfield{author}{\bibinfo{person}{Mariona Car{\'o}s}, \bibinfo{person}{Maite Garolera}, \bibinfo{person}{Petia Radeva}, {and} \bibinfo{person}{Xavier Giro-i Nieto}.} \bibinfo{year}{2020}\natexlab{}.
\newblock \showarticletitle{Automatic reminiscence therapy for dementia}. In \bibinfo{booktitle}{\emph{Proceedings of the 2020 International Conference on Multimedia Retrieval}}. \bibinfo{pages}{383--387}.
\newblock


\bibitem[Chamberlain et~al\mbox{.}(2012)]%
        {chamberlain2012research}
\bibfield{author}{\bibinfo{person}{Alan Chamberlain}, \bibinfo{person}{Andy Crabtree}, \bibinfo{person}{Tom Rodden}, \bibinfo{person}{Matt Jones}, {and} \bibinfo{person}{Yvonne Rogers}.} \bibinfo{year}{2012}\natexlab{}.
\newblock \showarticletitle{Research in the wild: understanding 'in the wild' approaches to design and development}. In \bibinfo{booktitle}{\emph{Proceedings of the Designing Interactive Systems Conference}}. \bibinfo{pages}{795--796}.
\newblock


\bibitem[Chattaraman et~al\mbox{.}(2019)]%
        {chattaraman2019should}
\bibfield{author}{\bibinfo{person}{Veena Chattaraman}, \bibinfo{person}{Wi-Suk Kwon}, \bibinfo{person}{Juan~E Gilbert}, {and} \bibinfo{person}{Kassandra Ross}.} \bibinfo{year}{2019}\natexlab{}.
\newblock \showarticletitle{Should AI-Based, conversational digital assistants employ social-or task-oriented interaction style? A task-competency and reciprocity perspective for older adults}.
\newblock \bibinfo{journal}{\emph{Computers in Human Behavior}}  \bibinfo{volume}{90} (\bibinfo{year}{2019}), \bibinfo{pages}{315--330}.
\newblock


\bibitem[Chu et~al\mbox{.}(2022)]%
        {chu2022digital}
\bibfield{author}{\bibinfo{person}{Charlene~H Chu}, \bibinfo{person}{Rune Nyrup}, \bibinfo{person}{Kathleen Leslie}, \bibinfo{person}{Jiamin Shi}, \bibinfo{person}{Andria Bianchi}, \bibinfo{person}{Alexandra Lyn}, \bibinfo{person}{Molly McNicholl}, \bibinfo{person}{Shehroz Khan}, \bibinfo{person}{Samira Rahimi}, {and} \bibinfo{person}{Amanda Grenier}.} \bibinfo{year}{2022}\natexlab{}.
\newblock \showarticletitle{Digital ageism: Challenges and opportunities in artificial intelligence for older adults}.
\newblock \bibinfo{journal}{\emph{The Gerontologist}} \bibinfo{volume}{62}, \bibinfo{number}{7} (\bibinfo{year}{2022}), \bibinfo{pages}{947--955}.
\newblock


\bibitem[Cotelli et~al\mbox{.}(2012)]%
        {cotelli2012reminiscence}
\bibfield{author}{\bibinfo{person}{Maria Cotelli}, \bibinfo{person}{Rosa Manenti}, {and} \bibinfo{person}{Orazio Zanetti}.} \bibinfo{year}{2012}\natexlab{}.
\newblock \showarticletitle{Reminiscence therapy in dementia: A review}.
\newblock \bibinfo{journal}{\emph{Maturitas}} \bibinfo{volume}{72}, \bibinfo{number}{3} (\bibinfo{year}{2012}), \bibinfo{pages}{203--205}.
\newblock


\bibitem[Czech et~al\mbox{.}(2020)]%
        {czech2020discovering}
\bibfield{author}{\bibinfo{person}{Elaine Czech}, \bibinfo{person}{Mina Shibasaki}, \bibinfo{person}{Keitaro Tsuchiya}, \bibinfo{person}{Roshan~L Peiris}, {and} \bibinfo{person}{Kouta Minamizawa}.} \bibinfo{year}{2020}\natexlab{}.
\newblock \showarticletitle{Discovering narratives: Multi-sensory approach towards designing with people with dementia}. In \bibinfo{booktitle}{\emph{Extended Abstracts of the 2020 CHI Conference on Human Factors in Computing Systems}}. \bibinfo{pages}{1--8}.
\newblock


\bibitem[Dassa(2018)]%
        {dassa2018musical}
\bibfield{author}{\bibinfo{person}{Ayelet Dassa}.} \bibinfo{year}{2018}\natexlab{}.
\newblock \showarticletitle{Musical Auto-Biography Interview (MABI) as promoting self-identity and well-being in the elderly through music and reminiscence}.
\newblock \bibinfo{journal}{\emph{Nordic Journal of Music Therapy}} \bibinfo{volume}{27}, \bibinfo{number}{5} (\bibinfo{year}{2018}), \bibinfo{pages}{419--430}.
\newblock


\bibitem[Dixon and Lazar(2020)]%
        {dixon2020approach}
\bibfield{author}{\bibinfo{person}{Emma Dixon} {and} \bibinfo{person}{Amanda Lazar}.} \bibinfo{year}{2020}\natexlab{}.
\newblock \showarticletitle{Approach matters: Linking practitioner approaches to technology design for people with dementia}. In \bibinfo{booktitle}{\emph{Proceedings of the 2020 CHI Conference on Human Factors in Computing Systems}}. \bibinfo{pages}{1--15}.
\newblock


\bibitem[Dosovitsky and Bunge(2021)]%
        {dosovitsky2021bonding}
\bibfield{author}{\bibinfo{person}{Gilly Dosovitsky} {and} \bibinfo{person}{Eduardo~L Bunge}.} \bibinfo{year}{2021}\natexlab{}.
\newblock \showarticletitle{Bonding with bot: user feedback on a chatbot for social isolation}.
\newblock \bibinfo{journal}{\emph{Frontiers in digital health}}  \bibinfo{volume}{3} (\bibinfo{year}{2021}), \bibinfo{pages}{735053}.
\newblock


\bibitem[Edmeads and Metatla(2019)]%
        {edmeads2019designing}
\bibfield{author}{\bibinfo{person}{James Edmeads} {and} \bibinfo{person}{Oussama Metatla}.} \bibinfo{year}{2019}\natexlab{}.
\newblock \showarticletitle{Designing for reminiscence with people with dementia}. In \bibinfo{booktitle}{\emph{Extended Abstracts of the 2019 CHI Conference on Human Factors in Computing Systems}}. \bibinfo{pages}{1--6}.
\newblock


\bibitem[Elias et~al\mbox{.}(2015)]%
        {elias2015effectiveness}
\bibfield{author}{\bibinfo{person}{Sharifah Munirah~Syed Elias}, \bibinfo{person}{Christine Neville}, {and} \bibinfo{person}{Theresa Scott}.} \bibinfo{year}{2015}\natexlab{}.
\newblock \showarticletitle{The effectiveness of group reminiscence therapy for loneliness, anxiety and depression in older adults in long-term care: A systematic review}.
\newblock \bibinfo{journal}{\emph{Geriatric nursing}} \bibinfo{volume}{36}, \bibinfo{number}{5} (\bibinfo{year}{2015}), \bibinfo{pages}{372--380}.
\newblock


\bibitem[Engelbrecht et~al\mbox{.}(2023)]%
        {engelbrecht2023music}
\bibfield{author}{\bibinfo{person}{Romy Engelbrecht}, \bibinfo{person}{Sunil Bhar}, {and} \bibinfo{person}{Joseph Ciorciari}.} \bibinfo{year}{2023}\natexlab{}.
\newblock \showarticletitle{Music-assisted reminiscence therapy with older adults: Feasibility, acceptability, and outcomes}.
\newblock \bibinfo{journal}{\emph{Music Therapy Perspectives}} \bibinfo{volume}{41}, \bibinfo{number}{1} (\bibinfo{year}{2023}), \bibinfo{pages}{37--46}.
\newblock


\bibitem[Feng et~al\mbox{.}(2021)]%
        {feng2021design}
\bibfield{author}{\bibinfo{person}{Yuanyuan Feng}, \bibinfo{person}{Yaxing Yao}, {and} \bibinfo{person}{Norman Sadeh}.} \bibinfo{year}{2021}\natexlab{}.
\newblock \showarticletitle{A design space for privacy choices: Towards meaningful privacy control in the internet of things}. In \bibinfo{booktitle}{\emph{Proceedings of the 2021 CHI Conference on Human Factors in Computing Systems}}. \bibinfo{pages}{1--16}.
\newblock


\bibitem[Gamborino et~al\mbox{.}(2021)]%
        {gamborino2021towards}
\bibfield{author}{\bibinfo{person}{Edwinn Gamborino}, \bibinfo{person}{Alberto Herrera~Ruiz}, \bibinfo{person}{Jing-Fen Wang}, \bibinfo{person}{Tsung-Yuan Tseng}, \bibinfo{person}{Su-Ling Yeh}, {and} \bibinfo{person}{Li-Chen Fu}.} \bibinfo{year}{2021}\natexlab{}.
\newblock \showarticletitle{Towards effective robot-assisted photo reminiscence: Personalizing interactions through visual understanding and inferring}. In \bibinfo{booktitle}{\emph{International Conference on Human-Computer Interaction}}. Springer, \bibinfo{pages}{335--349}.
\newblock


\bibitem[Garattini et~al\mbox{.}(2012)]%
        {garattini2012linking}
\bibfield{author}{\bibinfo{person}{Chiara Garattini}, \bibinfo{person}{Joseph Wherton}, {and} \bibinfo{person}{David Prendergast}.} \bibinfo{year}{2012}\natexlab{}.
\newblock \showarticletitle{Linking the lonely: an exploration of a communication technology designed to support social interaction among older adults}.
\newblock \bibinfo{journal}{\emph{Universal Access in the Information Society}} \bibinfo{volume}{11}, \bibinfo{number}{2} (\bibinfo{year}{2012}), \bibinfo{pages}{211--222}.
\newblock


\bibitem[Hallford and Mellor(2013)]%
        {hallford2013reminiscence}
\bibfield{author}{\bibinfo{person}{David Hallford} {and} \bibinfo{person}{David Mellor}.} \bibinfo{year}{2013}\natexlab{}.
\newblock \showarticletitle{Reminiscence-based therapies for depression: Should they be used only with older adults?}
\newblock \bibinfo{journal}{\emph{Clinical Psychology: Science and Practice}} \bibinfo{volume}{20}, \bibinfo{number}{4} (\bibinfo{year}{2013}), \bibinfo{pages}{452}.
\newblock


\bibitem[Han and Cai(2023)]%
        {han2023design}
\bibfield{author}{\bibinfo{person}{Ariel Han} {and} \bibinfo{person}{Zhenyao Cai}.} \bibinfo{year}{2023}\natexlab{}.
\newblock \showarticletitle{Design implications of generative AI systems for visual storytelling for young learners}. In \bibinfo{booktitle}{\emph{Proceedings of the 22nd Annual ACM Interaction Design and Children Conference}}. \bibinfo{pages}{470--474}.
\newblock


\bibitem[Hangal et~al\mbox{.}(2011)]%
        {hangal2011muse}
\bibfield{author}{\bibinfo{person}{Sudheendra Hangal}, \bibinfo{person}{Monica~S Lam}, {and} \bibinfo{person}{Jeffrey Heer}.} \bibinfo{year}{2011}\natexlab{}.
\newblock \showarticletitle{Muse: Reviving memories using email archives}. In \bibinfo{booktitle}{\emph{Proceedings of the 24th annual ACM symposium on User interface software and technology}}. \bibinfo{pages}{75--84}.
\newblock


\bibitem[Hedgeman et~al\mbox{.}(2018)]%
        {hedgeman2018perceived}
\bibfield{author}{\bibinfo{person}{Elizabeth Hedgeman}, \bibinfo{person}{Rebecca~E Hasson}, \bibinfo{person}{Carrie~A Karvonen-Gutierrez}, \bibinfo{person}{William~H Herman}, {and} \bibinfo{person}{Siob{\'a}n~D Harlow}.} \bibinfo{year}{2018}\natexlab{}.
\newblock \showarticletitle{Perceived stress across the midlife: longitudinal changes among a diverse sample of women, the Study of Women’s health Across the Nation (SWAN)}.
\newblock \bibinfo{journal}{\emph{Women's midlife health}}  \bibinfo{volume}{4} (\bibinfo{year}{2018}), \bibinfo{pages}{1--11}.
\newblock


\bibitem[Hoofnagle et~al\mbox{.}(2010)]%
        {hoofnagle2010different}
\bibfield{author}{\bibinfo{person}{Chris~Jay Hoofnagle}, \bibinfo{person}{Jennifer King}, \bibinfo{person}{Su Li}, {and} \bibinfo{person}{Joseph Turow}.} \bibinfo{year}{2010}\natexlab{}.
\newblock \showarticletitle{How different are young adults from older adults when it comes to information privacy attitudes and policies?}
\newblock \bibinfo{journal}{\emph{Available at SSRN 1589864}} (\bibinfo{year}{2010}).
\newblock


\bibitem[Horowitz(2004)]%
        {horowitz2004prevalence}
\bibfield{author}{\bibinfo{person}{Amy Horowitz}.} \bibinfo{year}{2004}\natexlab{}.
\newblock \showarticletitle{The prevalence and consequences of vision impairment in later life}.
\newblock \bibinfo{journal}{\emph{Topics in Geriatric Rehabilitation}} \bibinfo{volume}{20}, \bibinfo{number}{3} (\bibinfo{year}{2004}), \bibinfo{pages}{185--195}.
\newblock


\bibitem[Hsieh and Wang(2003)]%
        {hsieh2003effect}
\bibfield{author}{\bibinfo{person}{Hsiu-Fang Hsieh} {and} \bibinfo{person}{Jing-Jy Wang}.} \bibinfo{year}{2003}\natexlab{}.
\newblock \showarticletitle{Effect of reminiscence therapy on depression in older adults: a systematic review}.
\newblock \bibinfo{journal}{\emph{International journal of nursing studies}} \bibinfo{volume}{40}, \bibinfo{number}{4} (\bibinfo{year}{2003}), \bibinfo{pages}{335--345}.
\newblock


\bibitem[Hsieh et~al\mbox{.}(2011)]%
        {hsieh2011soundcapsule}
\bibfield{author}{\bibinfo{person}{Pei-Chen Hsieh}, \bibinfo{person}{RH Liang}, {and} \bibinfo{person}{HC Chen}.} \bibinfo{year}{2011}\natexlab{}.
\newblock \showarticletitle{SoundCapsule: The study of reminiscence triggered by utilizing sound media and technology}. In \bibinfo{booktitle}{\emph{Proceedings of the 4th world conference on design research}}, Vol.~\bibinfo{volume}{10}.
\newblock


\bibitem[Hughes et~al\mbox{.}(2021)]%
        {hughes2021generative}
\bibfield{author}{\bibinfo{person}{Rowan~T Hughes}, \bibinfo{person}{Liming Zhu}, {and} \bibinfo{person}{Tomasz Bednarz}.} \bibinfo{year}{2021}\natexlab{}.
\newblock \showarticletitle{Generative adversarial networks--enabled human--artificial intelligence collaborative applications for creative and design industries: A systematic review of current approaches and trends}.
\newblock \bibinfo{journal}{\emph{Frontiers in artificial intelligence}}  \bibinfo{volume}{4} (\bibinfo{year}{2021}), \bibinfo{pages}{604234}.
\newblock


\bibitem[Im et~al\mbox{.}(2023)]%
        {im2023less}
\bibfield{author}{\bibinfo{person}{Jane Im}, \bibinfo{person}{Ruiyi Wang}, \bibinfo{person}{Weikun Lyu}, \bibinfo{person}{Nick Cook}, \bibinfo{person}{Hana Habib}, \bibinfo{person}{Lorrie~Faith Cranor}, \bibinfo{person}{Nikola Banovic}, {and} \bibinfo{person}{Florian Schaub}.} \bibinfo{year}{2023}\natexlab{}.
\newblock \showarticletitle{Less is Not More: Improving Findability and Actionability of Privacy Controls for Online Behavioral Advertising}. In \bibinfo{booktitle}{\emph{Proceedings of the 2023 CHI Conference on Human Factors in Computing Systems}}. \bibinfo{pages}{1--33}.
\newblock


\bibitem[Istvandity(2017)]%
        {istvandity2017combining}
\bibfield{author}{\bibinfo{person}{Lauren Istvandity}.} \bibinfo{year}{2017}\natexlab{}.
\newblock \showarticletitle{Combining music and reminiscence therapy interventions for wellbeing in elderly populations: A systematic review}.
\newblock \bibinfo{journal}{\emph{Complementary therapies in clinical practice}}  \bibinfo{volume}{28} (\bibinfo{year}{2017}), \bibinfo{pages}{18--25}.
\newblock


\bibitem[Janata et~al\mbox{.}(2007)]%
        {janata2007characterisation}
\bibfield{author}{\bibinfo{person}{Petr Janata}, \bibinfo{person}{Stefan~T Tomic}, {and} \bibinfo{person}{Sonja~K Rakowski}.} \bibinfo{year}{2007}\natexlab{}.
\newblock \showarticletitle{Characterisation of music-evoked autobiographical memories}.
\newblock \bibinfo{journal}{\emph{Memory}} \bibinfo{volume}{15}, \bibinfo{number}{8} (\bibinfo{year}{2007}), \bibinfo{pages}{845--860}.
\newblock


\bibitem[Jayaratne(2016)]%
        {jayaratne2016memory}
\bibfield{author}{\bibinfo{person}{Keisha Jayaratne}.} \bibinfo{year}{2016}\natexlab{}.
\newblock \showarticletitle{The Memory Tree: Using Sound to Support Reminiscence}. In \bibinfo{booktitle}{\emph{Proceedings of the 2016 CHI Conference Extended Abstracts on Human Factors in Computing Systems}}. \bibinfo{pages}{116--121}.
\newblock


\bibitem[Jin et~al\mbox{.}(2023)]%
        {jin2023understanding}
\bibfield{author}{\bibinfo{person}{Yucheng Jin}, \bibinfo{person}{Wanling Cai}, \bibinfo{person}{Li Chen}, \bibinfo{person}{Yuwan Dai}, {and} \bibinfo{person}{Tonglin Jiang}.} \bibinfo{year}{2023}\natexlab{}.
\newblock \showarticletitle{Understanding Disclosure and Support for Youth Mental Health in Social Music Communities}.
\newblock \bibinfo{journal}{\emph{Proceedings of the ACM on Human-Computer Interaction}} \bibinfo{volume}{7}, \bibinfo{number}{CSCW1} (\bibinfo{year}{2023}), \bibinfo{pages}{1--32}.
\newblock


\bibitem[Lamme and Baars(1993)]%
        {lamme1993including}
\bibfield{author}{\bibinfo{person}{Simone Lamme} {and} \bibinfo{person}{Jan Baars}.} \bibinfo{year}{1993}\natexlab{}.
\newblock \showarticletitle{Including social factors in the analysis of reminiscence in elderly individuals}.
\newblock \bibinfo{journal}{\emph{The International Journal of Aging and Human Development}} \bibinfo{volume}{37}, \bibinfo{number}{4} (\bibinfo{year}{1993}), \bibinfo{pages}{297--311}.
\newblock


\bibitem[Lazar et~al\mbox{.}(2014)]%
        {lazar2014systematic}
\bibfield{author}{\bibinfo{person}{Amanda Lazar}, \bibinfo{person}{Hilaire Thompson}, {and} \bibinfo{person}{George Demiris}.} \bibinfo{year}{2014}\natexlab{}.
\newblock \showarticletitle{A systematic review of the use of technology for reminiscence therapy}.
\newblock \bibinfo{journal}{\emph{Health education \& behavior}} \bibinfo{volume}{41}, \bibinfo{number}{1\_suppl} (\bibinfo{year}{2014}), \bibinfo{pages}{51S--61S}.
\newblock


\bibitem[Lewis(1971)]%
        {lewis1971reminiscing}
\bibfield{author}{\bibinfo{person}{Charles~N Lewis}.} \bibinfo{year}{1971}\natexlab{}.
\newblock \showarticletitle{Reminiscing and self-concept in old age.}
\newblock \bibinfo{journal}{\emph{Journal of gerontology}} (\bibinfo{year}{1971}).
\newblock


\bibitem[Lin et~al\mbox{.}(2013)]%
        {lin2013hearing}
\bibfield{author}{\bibinfo{person}{Frank~R Lin}, \bibinfo{person}{Kristine Yaffe}, \bibinfo{person}{Jin Xia}, \bibinfo{person}{Qian-Li Xue}, \bibinfo{person}{Tamara~B Harris}, \bibinfo{person}{Elizabeth Purchase-Helzner}, \bibinfo{person}{Suzanne Satterfield}, \bibinfo{person}{Hilsa~N Ayonayon}, \bibinfo{person}{Luigi Ferrucci}, \bibinfo{person}{Eleanor~M Simonsick}, {et~al\mbox{.}}} \bibinfo{year}{2013}\natexlab{}.
\newblock \showarticletitle{Hearing loss and cognitive decline in older adults}.
\newblock \bibinfo{journal}{\emph{JAMA internal medicine}} \bibinfo{volume}{173}, \bibinfo{number}{4} (\bibinfo{year}{2013}), \bibinfo{pages}{293--299}.
\newblock


\bibitem[Lin et~al\mbox{.}(2003)]%
        {lin2003effect}
\bibfield{author}{\bibinfo{person}{Yen-Chun Lin}, \bibinfo{person}{Yu-Tzu Dai}, {and} \bibinfo{person}{Shiow-Li Hwang}.} \bibinfo{year}{2003}\natexlab{}.
\newblock \showarticletitle{The effect of reminiscence on the elderly population: a systematic review}.
\newblock \bibinfo{journal}{\emph{Public health nursing}} \bibinfo{volume}{20}, \bibinfo{number}{4} (\bibinfo{year}{2003}), \bibinfo{pages}{297--306}.
\newblock


\bibitem[Lin et~al\mbox{.}(2020)]%
        {lin2020caire}
\bibfield{author}{\bibinfo{person}{Zhaojiang Lin}, \bibinfo{person}{Peng Xu}, \bibinfo{person}{Genta~Indra Winata}, \bibinfo{person}{Farhad~Bin Siddique}, \bibinfo{person}{Zihan Liu}, \bibinfo{person}{Jamin Shin}, {and} \bibinfo{person}{Pascale Fung}.} \bibinfo{year}{2020}\natexlab{}.
\newblock \showarticletitle{Caire: An end-to-end empathetic chatbot}. In \bibinfo{booktitle}{\emph{Proceedings of the AAAI Conference on Artificial Intelligence}}, Vol.~\bibinfo{volume}{34}. \bibinfo{pages}{13622--13623}.
\newblock


\bibitem[Ma et~al\mbox{.}(2021)]%
        {ma2021one}
\bibfield{author}{\bibinfo{person}{Zhengyi Ma}, \bibinfo{person}{Zhicheng Dou}, \bibinfo{person}{Yutao Zhu}, \bibinfo{person}{Hanxun Zhong}, {and} \bibinfo{person}{Ji-Rong Wen}.} \bibinfo{year}{2021}\natexlab{}.
\newblock \showarticletitle{One chatbot per person: Creating personalized chatbots based on implicit user profiles}. In \bibinfo{booktitle}{\emph{Proceedings of the 44th international ACM SIGIR conference on research and development in information retrieval}}. \bibinfo{pages}{555--564}.
\newblock


\bibitem[Mackay and Fayard(1999)]%
        {mackay1999video}
\bibfield{author}{\bibinfo{person}{Wendy~E Mackay} {and} \bibinfo{person}{Anne~Laure Fayard}.} \bibinfo{year}{1999}\natexlab{}.
\newblock \showarticletitle{Video brainstorming and prototyping: techniques for participatory design}. In \bibinfo{booktitle}{\emph{CHI'99 extended abstracts on Human factors in computing systems}}. \bibinfo{pages}{118--119}.
\newblock


\bibitem[Matthew-Maich et~al\mbox{.}(2016)]%
        {matthew2016designing}
\bibfield{author}{\bibinfo{person}{Nancy Matthew-Maich}, \bibinfo{person}{Lauren Harris}, \bibinfo{person}{Jenny Ploeg}, \bibinfo{person}{Maureen Markle-Reid}, \bibinfo{person}{Ruta Valaitis}, \bibinfo{person}{Sarah Ibrahim}, \bibinfo{person}{Amiram Gafni}, \bibinfo{person}{Sandra Isaacs}, {et~al\mbox{.}}} \bibinfo{year}{2016}\natexlab{}.
\newblock \showarticletitle{Designing, implementing, and evaluating mobile health technologies for managing chronic conditions in older adults: a scoping review}.
\newblock \bibinfo{journal}{\emph{JMIR mHealth and uHealth}} \bibinfo{volume}{4}, \bibinfo{number}{2} (\bibinfo{year}{2016}), \bibinfo{pages}{e5127}.
\newblock


\bibitem[McGookin(2019)]%
        {mcgookin2019reveal}
\bibfield{author}{\bibinfo{person}{David McGookin}.} \bibinfo{year}{2019}\natexlab{}.
\newblock \showarticletitle{Reveal: investigating proactive location-based reminiscing with personal digital photo repositories}. In \bibinfo{booktitle}{\emph{Proceedings of the 2019 CHI Conference on Human Factors in Computing Systems}}. \bibinfo{pages}{1--14}.
\newblock


\bibitem[Mirowski et~al\mbox{.}(2023)]%
        {mirowski2023co}
\bibfield{author}{\bibinfo{person}{Piotr Mirowski}, \bibinfo{person}{Kory~W Mathewson}, \bibinfo{person}{Jaylen Pittman}, {and} \bibinfo{person}{Richard Evans}.} \bibinfo{year}{2023}\natexlab{}.
\newblock \showarticletitle{Co-Writing Screenplays and Theatre Scripts with Language Models: Evaluation by Industry Professionals}. In \bibinfo{booktitle}{\emph{Proceedings of the 2023 CHI Conference on Human Factors in Computing Systems}}. \bibinfo{pages}{1--34}.
\newblock


\bibitem[Molinari and Reichlin(1985)]%
        {molinari1985life}
\bibfield{author}{\bibinfo{person}{Victor Molinari} {and} \bibinfo{person}{Robert~E Reichlin}.} \bibinfo{year}{1985}\natexlab{}.
\newblock \showarticletitle{Life review reminiscence in the elderly: A review of the literature}.
\newblock \bibinfo{journal}{\emph{The International Journal of Aging and Human Development}} \bibinfo{volume}{20}, \bibinfo{number}{2} (\bibinfo{year}{1985}), \bibinfo{pages}{81--92}.
\newblock


\bibitem[Morrison et~al\mbox{.}(2021)]%
        {morrison2021older}
\bibfield{author}{\bibinfo{person}{Benjamin Morrison}, \bibinfo{person}{Lynne Coventry}, {and} \bibinfo{person}{Pam Briggs}.} \bibinfo{year}{2021}\natexlab{}.
\newblock \showarticletitle{How do Older Adults feel about engaging with Cyber-Security?}
\newblock \bibinfo{journal}{\emph{Human Behavior and Emerging Technologies}} \bibinfo{volume}{3}, \bibinfo{number}{5} (\bibinfo{year}{2021}), \bibinfo{pages}{1033--1049}.
\newblock


\bibitem[Narducci et~al\mbox{.}(2020)]%
        {narducci2020investigation}
\bibfield{author}{\bibinfo{person}{Fedelucio Narducci}, \bibinfo{person}{Pierpaolo Basile}, \bibinfo{person}{Marco de Gemmis}, \bibinfo{person}{Pasquale Lops}, {and} \bibinfo{person}{Giovanni Semeraro}.} \bibinfo{year}{2020}\natexlab{}.
\newblock \showarticletitle{An investigation on the user interaction modes of conversational recommender systems for the music domain}.
\newblock \bibinfo{journal}{\emph{User Modeling and User-Adapted Interaction}}  \bibinfo{volume}{30} (\bibinfo{year}{2020}), \bibinfo{pages}{251--284}.
\newblock


\bibitem[Nikitina et~al\mbox{.}(2018)]%
        {nikitina2018smart}
\bibfield{author}{\bibinfo{person}{Svetlana Nikitina}, \bibinfo{person}{Sara Callaioli}, {and} \bibinfo{person}{Marcos Baez}.} \bibinfo{year}{2018}\natexlab{}.
\newblock \showarticletitle{Smart conversational agents for reminiscence}. In \bibinfo{booktitle}{\emph{Proceedings of the 1st International Workshop on Software Engineering for Cognitive Services}}. \bibinfo{pages}{52--57}.
\newblock


\bibitem[Nova(2023)]%
        {nova2023generative}
\bibfield{author}{\bibinfo{person}{Kannan Nova}.} \bibinfo{year}{2023}\natexlab{}.
\newblock \showarticletitle{Generative AI in Healthcare: Advancements in Electronic Health Records, facilitating Medical Languages, and Personalized Patient Care}.
\newblock \bibinfo{journal}{\emph{Journal of Advanced Analytics in Healthcare Management}} \bibinfo{volume}{7}, \bibinfo{number}{1} (\bibinfo{year}{2023}), \bibinfo{pages}{115--131}.
\newblock


\bibitem[Odom et~al\mbox{.}(2019)]%
        {odom2019investigating}
\bibfield{author}{\bibinfo{person}{William Odom}, \bibinfo{person}{Ron Wakkary}, \bibinfo{person}{Jeroen Hol}, \bibinfo{person}{Bram Naus}, \bibinfo{person}{Pepijn Verburg}, \bibinfo{person}{Tal Amram}, {and} \bibinfo{person}{Amy Yo~Sue Chen}.} \bibinfo{year}{2019}\natexlab{}.
\newblock \showarticletitle{Investigating slowness as a frame to design longer-term experiences with personal data: A field study of olly}. In \bibinfo{booktitle}{\emph{Proceedings of the 2019 CHI Conference on Human Factors in Computing Systems}}. \bibinfo{pages}{1--16}.
\newblock


\bibitem[O'leary and Barry(1998)]%
        {o1998reminiscence}
\bibfield{author}{\bibinfo{person}{Eleanor O'leary} {and} \bibinfo{person}{Nicola Barry}.} \bibinfo{year}{1998}\natexlab{}.
\newblock \showarticletitle{Reminiscence therapy with older adults}.
\newblock \bibinfo{journal}{\emph{Journal of Social Work Practice}} \bibinfo{volume}{12}, \bibinfo{number}{2} (\bibinfo{year}{1998}), \bibinfo{pages}{159--165}.
\newblock


\bibitem[Park(2013)]%
        {park2013digital}
\bibfield{author}{\bibinfo{person}{Yong~Jin Park}.} \bibinfo{year}{2013}\natexlab{}.
\newblock \showarticletitle{Digital literacy and privacy behavior online}.
\newblock \bibinfo{journal}{\emph{Communication research}} \bibinfo{volume}{40}, \bibinfo{number}{2} (\bibinfo{year}{2013}), \bibinfo{pages}{215--236}.
\newblock


\bibitem[Peesapati et~al\mbox{.}(2010)]%
        {peesapati2010pensieve}
\bibfield{author}{\bibinfo{person}{S~Tejaswi Peesapati}, \bibinfo{person}{Victoria Schwanda}, \bibinfo{person}{Johnathon Schultz}, \bibinfo{person}{Matt Lepage}, \bibinfo{person}{So-yae Jeong}, {and} \bibinfo{person}{Dan Cosley}.} \bibinfo{year}{2010}\natexlab{}.
\newblock \showarticletitle{Pensieve: supporting everyday reminiscence}. In \bibinfo{booktitle}{\emph{Proceedings of the SIGCHI Conference on Human Factors in Computing Systems}}. \bibinfo{pages}{2027--2036}.
\newblock


\bibitem[Petrelli et~al\mbox{.}(2008)]%
        {petrelli2008autotopography}
\bibfield{author}{\bibinfo{person}{Daniela Petrelli}, \bibinfo{person}{Steve Whittaker}, {and} \bibinfo{person}{Jens Brockmeier}.} \bibinfo{year}{2008}\natexlab{}.
\newblock \showarticletitle{AutoTopography: what can physical mementos tell us about digital memories?}. In \bibinfo{booktitle}{\emph{Proceedings of the SIGCHI conference on Human Factors in computing systems}}. \bibinfo{pages}{53--62}.
\newblock


\bibitem[Pino et~al\mbox{.}(2015)]%
        {pino2015we}
\bibfield{author}{\bibinfo{person}{Maribel Pino}, \bibinfo{person}{M{\'e}lodie Boulay}, \bibinfo{person}{Fran{\c{c}}ois Jouen}, {and} \bibinfo{person}{Anne-Sophie Rigaud}.} \bibinfo{year}{2015}\natexlab{}.
\newblock \showarticletitle{“Are we ready for robots that care for us?” Attitudes and opinions of older adults toward socially assistive robots}.
\newblock \bibinfo{journal}{\emph{Frontiers in aging neuroscience}}  \bibinfo{volume}{7} (\bibinfo{year}{2015}), \bibinfo{pages}{141}.
\newblock


\bibitem[Rapp et~al\mbox{.}(2021)]%
        {rapp2021human}
\bibfield{author}{\bibinfo{person}{Amon Rapp}, \bibinfo{person}{Lorenzo Curti}, {and} \bibinfo{person}{Arianna Boldi}.} \bibinfo{year}{2021}\natexlab{}.
\newblock \showarticletitle{The human side of human-chatbot interaction: A systematic literature review of ten years of research on text-based chatbots}.
\newblock \bibinfo{journal}{\emph{International Journal of Human-Computer Studies}}  \bibinfo{volume}{151} (\bibinfo{year}{2021}), \bibinfo{pages}{102630}.
\newblock


\bibitem[Rhim et~al\mbox{.}(2022)]%
        {rhim2022application}
\bibfield{author}{\bibinfo{person}{Jungwook Rhim}, \bibinfo{person}{Minji Kwak}, \bibinfo{person}{Yeaeun Gong}, {and} \bibinfo{person}{Gahgene Gweon}.} \bibinfo{year}{2022}\natexlab{}.
\newblock \showarticletitle{Application of humanization to survey chatbots: Change in chatbot perception, interaction experience, and survey data quality}.
\newblock \bibinfo{journal}{\emph{Computers in Human Behavior}}  \bibinfo{volume}{126} (\bibinfo{year}{2022}), \bibinfo{pages}{107034}.
\newblock


\bibitem[Rogers and Mynatt(2003)]%
        {rogers2003can}
\bibfield{author}{\bibinfo{person}{Wendy~A Rogers} {and} \bibinfo{person}{Elizabeth~D Mynatt}.} \bibinfo{year}{2003}\natexlab{}.
\newblock \showarticletitle{How can technology contribute to the quality of life of older adults}.
\newblock \bibinfo{journal}{\emph{The technology of humanity: Can technology contribute to the quality of life}}  \bibinfo{volume}{22} (\bibinfo{year}{2003}), \bibinfo{pages}{30}.
\newblock


\bibitem[Romaniuk and Romaniuk(1981)]%
        {romaniuk1981looking}
\bibfield{author}{\bibinfo{person}{Michael Romaniuk} {and} \bibinfo{person}{Jean~Gasen Romaniuk}.} \bibinfo{year}{1981}\natexlab{}.
\newblock \showarticletitle{Looking back: An analysis of reminiscence functions and triggers}.
\newblock \bibinfo{journal}{\emph{Experimental aging research}} \bibinfo{volume}{7}, \bibinfo{number}{4} (\bibinfo{year}{1981}), \bibinfo{pages}{477--489}.
\newblock


\bibitem[Ryu et~al\mbox{.}(2020)]%
        {ryu2020simple}
\bibfield{author}{\bibinfo{person}{Hyeyoung Ryu}, \bibinfo{person}{Soyeon Kim}, \bibinfo{person}{Dain Kim}, \bibinfo{person}{Sooan Han}, \bibinfo{person}{Keeheon Lee}, {and} \bibinfo{person}{Younah Kang}.} \bibinfo{year}{2020}\natexlab{}.
\newblock \showarticletitle{Simple and steady interactions win the healthy mentality: Designing a chatbot service for the elderly}.
\newblock \bibinfo{journal}{\emph{Proceedings of the ACM on human-computer interaction}} \bibinfo{volume}{4}, \bibinfo{number}{CSCW2} (\bibinfo{year}{2020}), \bibinfo{pages}{1--25}.
\newblock


\bibitem[Sakaguchi-Tang et~al\mbox{.}(2021)]%
        {sakaguchi2021co}
\bibfield{author}{\bibinfo{person}{Dawn~K Sakaguchi-Tang}, \bibinfo{person}{Jay~L Cunningham}, \bibinfo{person}{Wendy Roldan}, \bibinfo{person}{Jason Yip}, {and} \bibinfo{person}{Julie~A Kientz}.} \bibinfo{year}{2021}\natexlab{}.
\newblock \showarticletitle{Co-design with older adults: examining and reflecting on collaboration with aging communities}.
\newblock \bibinfo{journal}{\emph{Proceedings of the ACM on Human-Computer Interaction}} \bibinfo{volume}{5}, \bibinfo{number}{CSCW2} (\bibinfo{year}{2021}), \bibinfo{pages}{1--28}.
\newblock


\bibitem[Sas et~al\mbox{.}(2020)]%
        {sas2020supporting}
\bibfield{author}{\bibinfo{person}{Corina Sas}, \bibinfo{person}{Nigel Davies}, \bibinfo{person}{Sarah Clinch}, \bibinfo{person}{Peter Shaw}, \bibinfo{person}{Mateusz Mikusz}, \bibinfo{person}{Madeleine Steeds}, {and} \bibinfo{person}{Lukas Nohrer}.} \bibinfo{year}{2020}\natexlab{}.
\newblock \showarticletitle{Supporting stimulation needs in dementia care through wall-sized displays}. In \bibinfo{booktitle}{\emph{Proceedings of the 2020 chi conference on human factors in computing systems}}. \bibinfo{pages}{1--16}.
\newblock


\bibitem[Sch{\"a}fer et~al\mbox{.}(2013)]%
        {schafer2013psychological}
\bibfield{author}{\bibinfo{person}{Thomas Sch{\"a}fer}, \bibinfo{person}{Peter Sedlmeier}, \bibinfo{person}{Christine St{\"a}dtler}, {and} \bibinfo{person}{David Huron}.} \bibinfo{year}{2013}\natexlab{}.
\newblock \showarticletitle{The psychological functions of music listening}.
\newblock \bibinfo{journal}{\emph{Frontiers in psychology}}  \bibinfo{volume}{4} (\bibinfo{year}{2013}), \bibinfo{pages}{511}.
\newblock


\bibitem[Schoenborn and Heyman(2009)]%
        {schoenborn2009health}
\bibfield{author}{\bibinfo{person}{Charlotte~A Schoenborn} {and} \bibinfo{person}{Kathleen~M Heyman}.} \bibinfo{year}{2009}\natexlab{}.
\newblock \showarticletitle{Health characteristics of adults aged 55 years and over: United States, 2004--2007}.
\newblock \bibinfo{journal}{\emph{Natl Health Stat Report}}  \bibinfo{volume}{16} (\bibinfo{year}{2009}), \bibinfo{pages}{1--31}.
\newblock


\bibitem[Schroeder et~al\mbox{.}(2022)]%
        {schroeder2022data}
\bibfield{author}{\bibinfo{person}{Tanja Schroeder}, \bibinfo{person}{Maximilian Haug}, {and} \bibinfo{person}{Heiko Gewald}.} \bibinfo{year}{2022}\natexlab{}.
\newblock \showarticletitle{Data privacy concerns using mhealth apps and smart speakers: Comparative interview study among mature adults}.
\newblock \bibinfo{journal}{\emph{JMIR formative research}} \bibinfo{volume}{6}, \bibinfo{number}{6} (\bibinfo{year}{2022}), \bibinfo{pages}{e28025}.
\newblock


\bibitem[Sedikides et~al\mbox{.}(2022)]%
        {sedikides2022psychological}
\bibfield{author}{\bibinfo{person}{Constantine Sedikides}, \bibinfo{person}{Joost Leunissen}, {and} \bibinfo{person}{Tim Wildschut}.} \bibinfo{year}{2022}\natexlab{}.
\newblock \showarticletitle{The psychological benefits of music-evoked nostalgia}.
\newblock \bibinfo{journal}{\emph{Psychology of Music}} \bibinfo{volume}{50}, \bibinfo{number}{6} (\bibinfo{year}{2022}), \bibinfo{pages}{2044--2062}.
\newblock


\bibitem[Selwyn et~al\mbox{.}(2003)]%
        {selwyn2003older}
\bibfield{author}{\bibinfo{person}{Neil Selwyn}, \bibinfo{person}{Stephen Gorard}, \bibinfo{person}{John Furlong}, {and} \bibinfo{person}{Louise Madden}.} \bibinfo{year}{2003}\natexlab{}.
\newblock \showarticletitle{Older adults' use of information and communications technology in everyday life}.
\newblock \bibinfo{journal}{\emph{Ageing \& Society}} \bibinfo{volume}{23}, \bibinfo{number}{5} (\bibinfo{year}{2003}), \bibinfo{pages}{561--582}.
\newblock


\bibitem[Smyth(1998)]%
        {smyth1998written}
\bibfield{author}{\bibinfo{person}{Joshua~M Smyth}.} \bibinfo{year}{1998}\natexlab{}.
\newblock \showarticletitle{Written emotional expression: effect sizes, outcome types, and moderating variables.}
\newblock \bibinfo{journal}{\emph{Journal of consulting and clinical psychology}} \bibinfo{volume}{66}, \bibinfo{number}{1} (\bibinfo{year}{1998}), \bibinfo{pages}{174}.
\newblock


\bibitem[Srinivasan and Uchino(2021)]%
        {srinivasan2021biases}
\bibfield{author}{\bibinfo{person}{Ramya Srinivasan} {and} \bibinfo{person}{Kanji Uchino}.} \bibinfo{year}{2021}\natexlab{}.
\newblock \showarticletitle{Biases in generative art: A causal look from the lens of art history}. In \bibinfo{booktitle}{\emph{Proceedings of the 2021 ACM Conference on Fairness, Accountability, and Transparency}}. \bibinfo{pages}{41--51}.
\newblock


\bibitem[Stigall et~al\mbox{.}(2019)]%
        {stigall2019older}
\bibfield{author}{\bibinfo{person}{Brodrick Stigall}, \bibinfo{person}{Jenny Waycott}, \bibinfo{person}{Steven Baker}, {and} \bibinfo{person}{Kelly Caine}.} \bibinfo{year}{2019}\natexlab{}.
\newblock \showarticletitle{Older adults' perception and use of voice user interfaces: a preliminary review of the computing literature}. In \bibinfo{booktitle}{\emph{Proceedings of the 31st Australian Conference on Human-Computer-Interaction}}. \bibinfo{pages}{423--427}.
\newblock


\bibitem[Tsai et~al\mbox{.}(2013)]%
        {tsai2013framing}
\bibfield{author}{\bibinfo{person}{Wenn-Chieh Tsai}, \bibinfo{person}{Hung-Chi Lee}, \bibinfo{person}{Joey Chiao-Yin Hsiao}, \bibinfo{person}{Rung-Huei Liang}, {and} \bibinfo{person}{Jane Yung-jen Hsu}.} \bibinfo{year}{2013}\natexlab{}.
\newblock \showarticletitle{Framing design of reminiscence aids with transactive memory theory}.
\newblock In \bibinfo{booktitle}{\emph{CHI'13 Extended Abstracts on Human Factors in Computing Systems}}. \bibinfo{pages}{331--336}.
\newblock


\bibitem[Vert(2023)]%
        {vert2023will}
\bibfield{author}{\bibinfo{person}{Jean-Philippe Vert}.} \bibinfo{year}{2023}\natexlab{}.
\newblock \showarticletitle{How will generative AI disrupt data science in drug discovery?}
\newblock \bibinfo{journal}{\emph{Nature Biotechnology}} (\bibinfo{year}{2023}), \bibinfo{pages}{1--2}.
\newblock


\bibitem[Wang(2007)]%
        {wang2007group}
\bibfield{author}{\bibinfo{person}{Jing-Jy Wang}.} \bibinfo{year}{2007}\natexlab{}.
\newblock \showarticletitle{Group reminiscence therapy for cognitive and affective function of demented elderly in Taiwan}.
\newblock \bibinfo{journal}{\emph{International Journal of Geriatric Psychiatry: A journal of the psychiatry of late life and allied sciences}} \bibinfo{volume}{22}, \bibinfo{number}{12} (\bibinfo{year}{2007}), \bibinfo{pages}{1235--1240}.
\newblock


\bibitem[Wang et~al\mbox{.}(2019)]%
        {wang2019technology}
\bibfield{author}{\bibinfo{person}{Shengzhi Wang}, \bibinfo{person}{Khalisa Bolling}, \bibinfo{person}{Wenlin Mao}, \bibinfo{person}{Jennifer Reichstadt}, \bibinfo{person}{Dilip Jeste}, \bibinfo{person}{Ho-Cheol Kim}, {and} \bibinfo{person}{Camille Nebeker}.} \bibinfo{year}{2019}\natexlab{}.
\newblock \showarticletitle{Technology to support aging in place: Older adults’ perspectives}. In \bibinfo{booktitle}{\emph{Healthcare}}, Vol.~\bibinfo{volume}{7}. MDPI, \bibinfo{pages}{60}.
\newblock


\bibitem[Wang and Yin(2021)]%
        {wang2021explanations}
\bibfield{author}{\bibinfo{person}{Xinru Wang} {and} \bibinfo{person}{Ming Yin}.} \bibinfo{year}{2021}\natexlab{}.
\newblock \showarticletitle{Are explanations helpful? a comparative study of the effects of explanations in ai-assisted decision-making}. In \bibinfo{booktitle}{\emph{26th international conference on intelligent user interfaces}}. \bibinfo{pages}{318--328}.
\newblock


\bibitem[Weisz et~al\mbox{.}(2023)]%
        {weisz2023toward}
\bibfield{author}{\bibinfo{person}{Justin~D. Weisz}, \bibinfo{person}{Michael~J. Muller}, \bibinfo{person}{Jessica He}, {and} \bibinfo{person}{Stephanie Houde}.} \bibinfo{year}{2023}\natexlab{}.
\newblock \showarticletitle{Toward General Design Principles for Generative {AI} Applications 130-144}. In \bibinfo{booktitle}{\emph{Joint Proceedings of the {IUI} 2023 Workshops: HAI-GEN, ITAH, MILC, SHAI, SketchRec, {SOCIALIZE} co-located with the {ACM} International Conference on Intelligent User Interfaces {(IUI} 2023), Sydney, Australia, March 27-31, 2023}} \emph{(\bibinfo{series}{{CEUR} Workshop Proceedings}, Vol.~\bibinfo{volume}{3359})}. \bibinfo{publisher}{CEUR-WS.org}, \bibinfo{pages}{130--144}.
\newblock


\bibitem[Westerhof and Bohlmeijer(2014)]%
        {westerhof2014celebrating}
\bibfield{author}{\bibinfo{person}{Gerben~J Westerhof} {and} \bibinfo{person}{Ernst~T Bohlmeijer}.} \bibinfo{year}{2014}\natexlab{}.
\newblock \showarticletitle{Celebrating fifty years of research and applications in reminiscence and life review: State of the art and new directions}.
\newblock \bibinfo{journal}{\emph{Journal of Aging studies}}  \bibinfo{volume}{29} (\bibinfo{year}{2014}), \bibinfo{pages}{107--114}.
\newblock


\bibitem[Wong(2013)]%
        {wong2013processes}
\bibfield{author}{\bibinfo{person}{Paul~TP Wong}.} \bibinfo{year}{2013}\natexlab{}.
\newblock \showarticletitle{The processes of adaptive reminiscence}.
\newblock In \bibinfo{booktitle}{\emph{The art and science of reminiscing}}. \bibinfo{publisher}{Taylor \& Francis}, \bibinfo{pages}{23--35}.
\newblock


\bibitem[Woods et~al\mbox{.}(1992)]%
        {woods1992reminiscence}
\bibfield{author}{\bibinfo{person}{Bob Woods}, \bibinfo{person}{Sara Portnoy}, \bibinfo{person}{Donna Head}, {and} \bibinfo{person}{Gemma Jones}.} \bibinfo{year}{1992}\natexlab{}.
\newblock \showarticletitle{Reminiscence and life review with persons with dementia: which way forward?}
\newblock In \bibinfo{booktitle}{\emph{Care-giving in Dementia}}. \bibinfo{publisher}{Routledge}, \bibinfo{pages}{137--161}.
\newblock


\bibitem[Yoo et~al\mbox{.}(2020)]%
        {yoo2020understanding}
\bibfield{author}{\bibinfo{person}{MinYoung Yoo}, \bibinfo{person}{William Odom}, {and} \bibinfo{person}{Arne Berger}.} \bibinfo{year}{2020}\natexlab{}.
\newblock \showarticletitle{Understanding how audio mediates experiences of reminiscence for people living with blindness}. In \bibinfo{booktitle}{\emph{Companion Publication of the 2020 ACM Designing Interactive Systems Conference}}. \bibinfo{pages}{73--78}.
\newblock


\bibitem[Yoo et~al\mbox{.}(2021)]%
        {yoo2021understanding}
\bibfield{author}{\bibinfo{person}{MinYoung Yoo}, \bibinfo{person}{William Odom}, {and} \bibinfo{person}{Arne Berger}.} \bibinfo{year}{2021}\natexlab{}.
\newblock \showarticletitle{Understanding everyday experiences of reminiscence for people with blindness: Practices, tensions and probing new design possibilities}. In \bibinfo{booktitle}{\emph{Proceedings of the 2021 CHI Conference on Human Factors in Computing Systems}}. \bibinfo{pages}{1--15}.
\newblock


\bibitem[Yusif et~al\mbox{.}(2016)]%
        {yusif2016older}
\bibfield{author}{\bibinfo{person}{Salifu Yusif}, \bibinfo{person}{Jeffrey Soar}, {and} \bibinfo{person}{Abdul Hafeez-Baig}.} \bibinfo{year}{2016}\natexlab{}.
\newblock \showarticletitle{Older people, assistive technologies, and the barriers to adoption: A systematic review}.
\newblock \bibinfo{journal}{\emph{International journal of medical informatics}}  \bibinfo{volume}{94} (\bibinfo{year}{2016}), \bibinfo{pages}{112--116}.
\newblock


\bibitem[Zhang et~al\mbox{.}(2023)]%
        {zhang2023complete}
\bibfield{author}{\bibinfo{person}{Chaoning Zhang}, \bibinfo{person}{Chenshuang Zhang}, \bibinfo{person}{Sheng Zheng}, \bibinfo{person}{Yu Qiao}, \bibinfo{person}{Chenghao Li}, \bibinfo{person}{Mengchun Zhang}, \bibinfo{person}{Sumit~Kumar Dam}, \bibinfo{person}{Chu~Myaet Thwal}, \bibinfo{person}{Ye~Lin Tun}, \bibinfo{person}{Le~Luang Huy}, {et~al\mbox{.}}} \bibinfo{year}{2023}\natexlab{}.
\newblock \showarticletitle{A complete survey on generative ai (aigc): Is chatgpt from gpt-4 to gpt-5 all you need?}
\newblock \bibinfo{journal}{\emph{arXiv preprint arXiv:2303.11717}} (\bibinfo{year}{2023}).
\newblock


\bibitem[Zwinderman et~al\mbox{.}(2013)]%
        {zwinderman2013using}
\bibfield{author}{\bibinfo{person}{Matthijs Zwinderman}, \bibinfo{person}{Rinze Leenheer}, \bibinfo{person}{Azadeh Shirzad}, \bibinfo{person}{Nikolay Chupriyanov}, \bibinfo{person}{Glenn Veugen}, \bibinfo{person}{Biyong Zhang}, {and} \bibinfo{person}{Panos Markopoulos}.} \bibinfo{year}{2013}\natexlab{}.
\newblock \showarticletitle{Using video prototypes for evaluating design concepts with users: a comparison to usability testing}. In \bibinfo{booktitle}{\emph{Human-Computer Interaction--INTERACT 2013: 14th IFIP TC 13 International Conference, Cape Town, South Africa, September 2-6, 2013, Proceedings, Part II 14}}. Springer, \bibinfo{pages}{774--781}.
\newblock


\end{thebibliography}

\end{document}